\documentclass[12pt]{article}

\usepackage{latexsym,amsmath,amssymb,amsthm,amsfonts,graphicx,natbib,url}
\usepackage{epsfig}
\usepackage{setspace}
\usepackage{bm}
\usepackage[textwidth=6in,textheight=8.5in,centering]{geometry}

\newcommand*\patchAmsMathEnvironmentForLineno[1]{%
      \expandafter\let\csname old#1\expandafter\endcsname\csname #1\endcsname
      \expandafter\let\csname oldend#1\expandafter\endcsname\csname end#1\endcsname
      \renewenvironment{#1}%
         {\linenomath\csname old#1\endcsname}%
         {\csname oldend#1\endcsname\endlinenomath}}%
    \newcommand*\patchBothAmsMathEnvironmentsForLineno[1]{%
      \patchAmsMathEnvironmentForLineno{#1}%
      \patchAmsMathEnvironmentForLineno{#1*}}%
    \AtBeginDocument{%
    \patchBothAmsMathEnvironmentsForLineno{equation}%
    \patchBothAmsMathEnvironmentsForLineno{align}%
    \patchBothAmsMathEnvironmentsForLineno{flalign}%
    \patchBothAmsMathEnvironmentsForLineno{alignat}%
    \patchBothAmsMathEnvironmentsForLineno{gather}%
    \patchBothAmsMathEnvironmentsForLineno{multline}%
    }
\usepackage[mathlines,displaymath]{lineno}	
\usepackage[usenames, dvipsnames]{color}

\makeatletter

\include{def}
\def\dispmuskip{\thinmuskip= 3mu plus 0mu minus 2mu \medmuskip=  4mu plus 2mu minus 2mu \thickmuskip=5mu plus 5mu minus 2mu}
\def\textmuskip{\thinmuskip= 0mu                    \medmuskip=  1mu plus 1mu minus 1mu \thickmuskip=2mu plus 3mu minus 1mu}
\def\beq{\dispmuskip\begin{equation}}    \def\eeq{\end{equation}\textmuskip}
\def\beqn{\dispmuskip\begin{displaymath}}\def\eeqn{\end{displaymath}\textmuskip}
\def\bea{\dispmuskip\begin{eqnarray}}    \def\eea{\end{eqnarray}\textmuskip}
\def\bean{\dispmuskip\begin{eqnarray*}}  \def\eean{\end{eqnarray*}\textmuskip}

\def\paradot#1{\vspace{1.3ex plus 0.7ex minus 0.5ex}\noindent{\bf\boldmath{#1.}}}

\newtheorem{algorithm}{Algorithm}

\newcommand{\KL}{Kullback-Leibler}

\newcommand{\diag}{\text{diag}}

\newcommand{\eps}{\epsilon}
\newcommand{\wh}{\widehat}
\newcommand{\wt}{\widetilde}

\def\v{\boldsymbol}

\def\E{{\mathbb E}}                         

\def\P{{\rm P}}                         

\def\v{\boldsymbol}

\def\a{\alpha}

\def\t{\theta}
\def\b{\beta}
\def\l{\lambda}

\def\N{{\cal N}}

\def\Kl{\text{\rm KL}}
\def\LB{\text{\rm LB}}

\def\vec{\text{\rm vec}}

\def\cov{\text{\rm cov}}

\def\argmax{\text{\rm argmax}}

\def\diag{\text{\rm diag}}

\makeatother

\begin{document}
\title{Bayesian Deep Net GLM and GLMM}
\author{M.-N. Tran\thanks{\textit{Discipline of Business Analytics, The University of Sydney Business School and ACEMS.}}
\and N. Nguyen\footnotemark[1]
\and D. Nott \thanks{\textit{Department of Statistics and Applied Probability, National University of Singapore.}}
\and R. Kohn\thanks{\textit{School of Economics, UNSW Business School and ACEMS.}}
}
\date{\empty}
\maketitle
\begin{abstract}
Deep feedforward neural networks (DFNNs) are a powerful tool for functional approximation.
We describe flexible versions of generalized linear and generalized linear
mixed models incorporating basis functions formed by a DFNN.
The consideration of neural networks with random effects is not widely used in the literature, perhaps
because of the computational challenges of incorporating subject specific parameters into already complex models.  
Efficient computational methods for high-dimensional Bayesian inference are developed using Gaussian variational approximation,  
with a parsimonious but flexible factor parametrization of the covariance matrix. 
We implement natural gradient methods for the optimization, 
exploiting the factor structure of the variational covariance matrix in computation of the natural gradient.
Our flexible DFNN models and Bayesian inference approach lead to a regression and classification method that has a high prediction accuracy, and is able to quantify the prediction uncertainty in a principled and convenient way. We also describe how to perform variable selection in our deep learning method. The proposed methods are illustrated in a wide range of simulated and real-data examples, and the results compare favourably to a state of the art flexible regression and classification method in the statistical literature, the Bayesian additive regression trees (BART) method. User-friendly software packages in Matlab, R and Python implementing the proposed methods are available at \texttt{https://github.com/VBayesLab}.

\paradot{Keywords} Deep learning; Factor models; Reparametrization gradient; Stochastic optimization; Variational approximation; Variable selection.

\end{abstract}

\section{Introduction}\label{sec:Introduction}
Deep feedforward neural network (DFNN) modeling provides a powerful technique for approximating regression functions with multiple regressors,
and has become increasingly popular recently.  
DFNNs have been applied successfully in fields such as image processing, computer vision and language recognition.
See \cite{Schmidhuber:2015} for a historical survey and \cite{Goodfellow:2016} for a more comprehensive recent discussion of DFNNs and other types of neural networks, collectively known as deep learning models. 

This paper considers variants of generalized linear models (GLMs) and generalized linear mixed model (GLMMs) using DFNNs as a way to efficiently transform
a vector of $p$ raw covariates $X=(X_1,...,X_p)^\top$ into a new vector of $m$ predictors $Z$ in the model.
We refer to these DFNN-based versions of GLM and GLMM as DeepGLM and DeepGLMM, respectively.  
A conventional GLM uses a link function that links the conditional mean of the response variable $Y$ 
to a linear combination of the predictors $Z=\phi(X)=(\phi_1(X),...,\phi_m(X))^\top$, with each $\phi_j(X)$ a function of $X$.
We refer to the original raw input variables $X_j$ as covariates, and refer to the transformations $\phi_j(X)$ as predictors.
In conventional GLMs, the $\phi_j(X)$ are chosen {\it a priori} in some way before any model selection, 
but here we are concerned with learning an appropriate $Z$ from data through a flexible smooth transformation.
In the machine learning literature the predictors $Z$ are commonly referred to as learned features.
If a DFNN is used for transforming the covariates, then $Z$ has the form
\beq\label{eq:NN1}
Z=f_L\Big(W_L,f_{L-1}\big(W_{L-1},\cdots f_1(W_1,X)\cdots\big)\Big),
\eeq
\begin{figure}[ht]
\centering
\includegraphics[width=150mm]{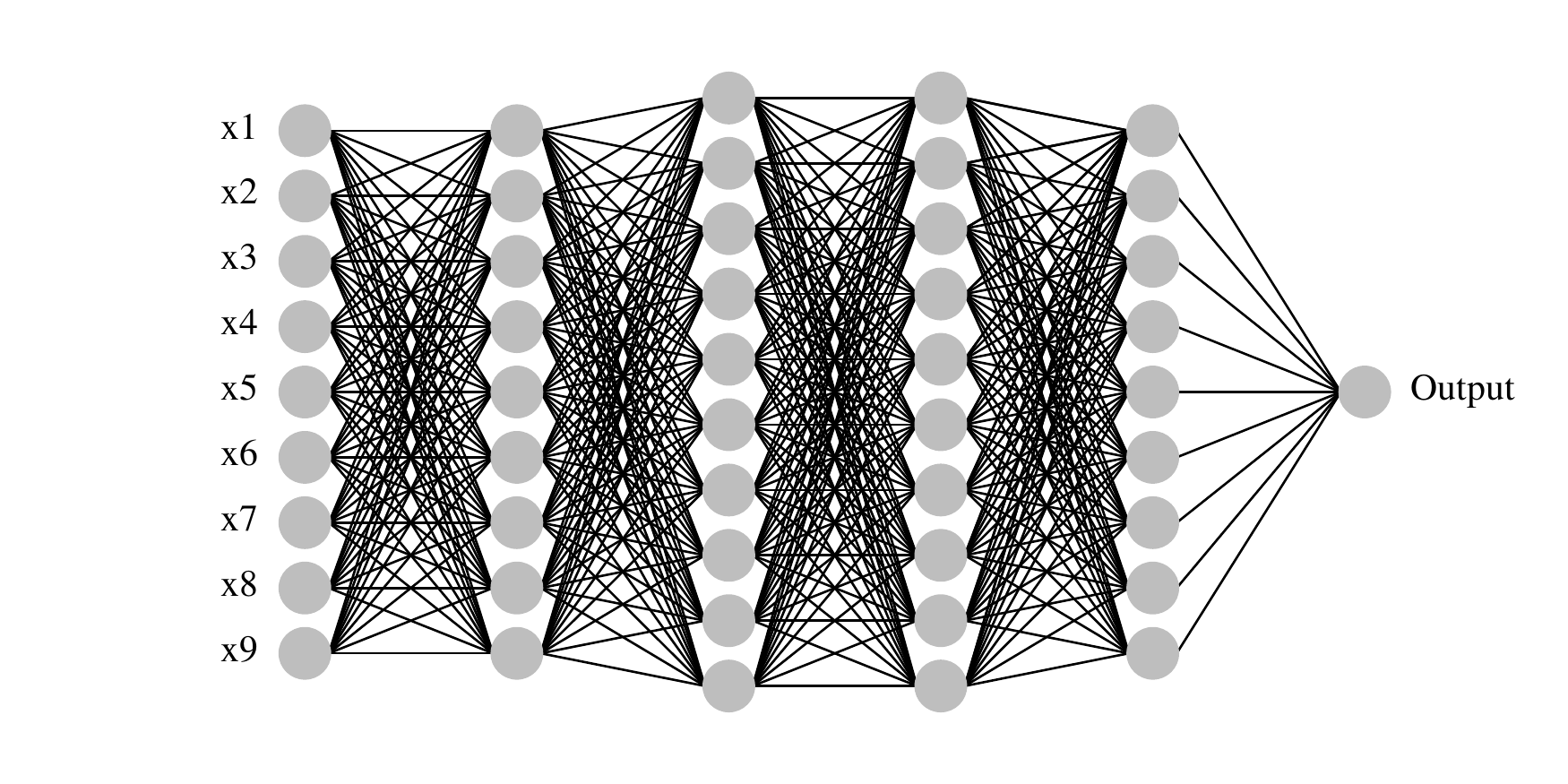}
\caption{Graphical representation of a layered composition function with $L=4$ hidden layers. The input layer represents 9 raw covariates $X$. The last hidden layer (hidden layer 4) represents the predictors $Z$.}
\label{f:neural net}
\end{figure}
which is often graphically represented by a network as in Figure \ref{f:neural net}.
Each vector-valued function $Z_l=f_l(W_l,Z_{l-1})$, $l=1,2,\dots,L$, is called a hidden layer, $L$ is the number of hidden layers in the network, $w=(W_1,...,W_L)$ is the set of weights, and we define $Z_0=X$.  The function 
$f_l(W_l,Z_{l-1})$ is assumed to be of the form 
$h_l(W_lZ_{l-1})$, where $W_l$ is a matrix of weights that connect layer $l$ to layer $l+1$ and $h_{l}(\cdot)$ is a scalar function, called the activation function. 
Applying $h_l$ to a vector should be understood component-wise.  For a discussion of alternative kinds of architectures for deep neural networks see, for example, \cite{Goodfellow:2016}.  
The architecture \eqref{eq:NN1} provides a powerful way to transform 
the raw covariate data $X$ into summary statistics $Z$ that have some desirable properties.  
In the DeepGLM, we link the conditional mean of the response $\xi=\E(Y|X)$ to a linear combination of $Z$
\[g(\xi)=\beta_0+ Z^\top\wt\beta.\]
and the model parameters consist of $w$, $\beta$ and other possible parameters such as any dispersion parameters.  
Section \ref{sec:DeepGLMM} defines DeepGLMM models by adding random effects to DeepGLM models. Such models are designed to model within-subject dependence.
The use of neural network basis functions in the context of mixed effects models seems not considered
much in the literature. 
\citet{Lai2006} is the only work that we know of that deals with a neural network basis and random effects, and they use it for modeling pharmacokinetic data.
They consider neural networks with only one hidden layer, however.

Estimation and variable selection in complex and high-dimensional models like DeepGLM and DeepGLMM are challenging.
This article develops Bayesian inference based on variational approximation,
which provides an approach to approximate Bayesian inference that is useful for many modern applications involving 
complex models and large datasets.  
We also consider variable selection, which is often of primary interest in statistics but is somewhat overlooked in the deep learning literature.
We describe a Bayesian adaptive group lasso method \citep{Leng:2013,Kyung:etal:2010}
 for variable selection in DeepGLM and DeepGLMM, in which the adaptive shrinkage parameters are estimated using marginal likelihood maximization 
with the optimization updating procedure conveniently embeded within the variational approximation.
Our variational approximation scheme assumes a multivariate Gaussian approximating family.  
With such a family, parsimonious but flexible methods for parametrizing the covariance matrix are necessary
if the approach is to be useful for problems with a high-dimensional parameter.  
Here we consider factor parametrizations \citep{Bartholomew2011} which are often effective for describing dependence in
high dimensional settings.  We discuss efficient methods for performing the variational optimization in this context using the natural gradient \citep{Amari:1998} by leveraging
the factor structure and using iterative conjugate gradient methods for solving large linear systems.  
In the case with a one-factor decomposition, 
we show that the natural gradient can be computed efficiently without iterative conjugate gradient methods,
which leads to a particularly simple Gaussian variational method for fitting high-dimensional models that can often be adequate
for predictive inference. We will refer to this estimation method as the NAtural gradient Gaussian Variational Approximation with factor Covariance (NAGVAC).

We illustrate the DeepGLM and DeepGLMM
and the training method NAGVAC in a range of experimental studies and applications, with a focus on datasets of only moderate size where quantification of prediction uncertainty is important.  
The Bayesian approach we follow is attractive in these applications
as it allows a principled and automatic way for selecting the ({\it many}) shrinkage parameters; see Section \ref{Variable selection}.
The Bayesian approach also leads to a principled and convenient way for quantifying prediction uncertainty through prediction intervals on test data,
which would be challenging to do so with non-Bayesian approaches for large models like DeepGLM.    
Many successful applications of deep learning are in image processing and speech recognition, where the datasets are large and have some special domain-application characteristics
such as association between the local pixels in an image. Skepticism has sometimes been expressed about whether deep learning methods are useful for applications
involving limited data where both prediction accuracy and quantifying prediction uncertainty are the focus. 
The main conclusion in our examples is that the DFNN-based regression models DeepGLM and DeepGLMM perform very well in terms of prediction accuracy and prediction uncertainty,
provided appropriate attention is paid to regularization methods, prior distributions and computational algorithms.  
We obtain results which compare favourably to the Bayesian Additive Regression Trees method (BART) of \cite{chipman2010}, which is a commonly-used flexible regression and classification method in the statistics literature.  
Software packages in Matlab, R and Python implementing the proposed methods are available at \texttt{https://github.com/vbayeslab}.

\paradot{Related work} \cite{Polson:2017} provide a Bayesian perspective on DFNN methodology and explain its interesting connection to statistical techniques such as 
principal component analysis and reduced rank regression. However, they do not provide any training method, which is a challenging problem
in Bayesian inference with DFNN.

The NAGVAC method is closely related to \cite{ong+ns16}, who consider Gaussian variational approximation with a factor covariance structure using stochastic gradient approaches for the
optimization.  Their approach uses the so-called reparametrization trick \citep{Kingma:2013,rezende+mw14} and its modification by \citet{Roeder2017} for estimating
gradients of the variational objective.  However, for certain models where it is very challenging to optimize the variational objective, first-order optimization methods
such as those considered in \cite{ong+ns16} may be very slow to converge.  \citet{ong+ntsd17} consider natural gradient methods for Gaussian
variational approximations using a factor covariance structure. However, their work was in the context of likelihood-free inference methods with a parameter
of dimension at most a few hundred, and the approach they develop does not scale to larger problems.   
The current work shows how the natural gradient Gaussian variational approximation with factor covariance can be implemented in very high dimensions.

Much of the recent literature relevant to our training method occurs in the field of deep learning.  
\cite{Martens:2010} considers second-order optimization methods in deep learning and describes Hessian-free optimization methods, adapting a 
long history of related methods in the numerical analysis literature to that context.  A detailed discussion of the connections between
natural gradient methods and second order optimization methods is given in \cite{Martens:2014}.  \cite{Pascanu:2014} consider the 
use of the natural gradient in deep learning problems and its connections with Hessian-free optimization, but like \cite{Martens:2010} their work is not specifically concerned with
variational objectives.  \cite{fan+wbkh15} consider how to implement
Hessian-free methods for the case of a variational objective function and Gaussian approximating family.  Similarly to our development here, 
they consider using conjugate gradient linear solvers as an efficient solution to the difficult matrix calculations that occur in a naive formulation
of second-order methods.  By a reparametrization approach they are able to obtain estimates of Hessian vector products.  They do not consider
factor parametrizations of the covariance structure, however.  Here we leverage the factor covariance structure to calculate the matrix vector products
we need directly, without the need to store large matrices.
Recently \cite{regier+ja17} consider a second
order trust region method for black box variational inference.  They show that while their approach may be more expensive per iteration than common
first-order methods for variational Gaussian approximation such as those implemented in \citet{titsias+l14} and \citet{Kucukelbir2016}, it reduces total computation
time and provably converges to a stationary point.  It may be useful to consider how to exploit a factor parametrization of the covariance
structure in the framework they develop, but this is not considered here.

The next section describes the two new classes of flexible statistical models DeepGLM and DeepGLMM.
Section \ref{sec:SVB} describes the natural gradient Gaussian variational approximation method.
Section \ref{Variable selection} describes the variable selection method and highlights advantages of our fully Bayesian treatment.
Section \ref{Bayesian treament} presents practical recommendations in training.
Section \ref{Applications} presents experimental studies and applications, and Section \ref{sec:conclusions} concludes.  
The Appendix gives details of the natural gradient computation and more details of an example.

\section{Flexible regression models with DFNN}\label{Flexible regression models with DFNN}
This section presents the DeepGLM and DeepGLMM models.
Deep feedforward neural network models, or multi-layer perceptrons, for classification and regression with a continuous response  
have been widely used in the machine learning literature.
We use statistical terminologies and unify these models under the popular GLM framework,
and propose the DFNN-based version of GLMM for analyzing panel data.

\subsection{DeepGLM}\label{sec:DeepGLM}
Consider a dataset $D=\{(y_{i},x_{i}),i=1,...,n\}$ with $y_{i}$ the response
and $x_{i}=(x_{i1},...,x_{ip})^\top$ the vector of $p$ covariates. 
We also use $y$ and $x$ to denote a generic response and a covariate vector, respectively.
Consider a neural net with the input vector $x$
and a scalar output as represented in Figure \ref{f:neural net}.
Denote by $z_j=\phi_j(x,w)$, $j=1,...,m$, the units in the last hidden layer, where $w$ is the weights up to the last hidden layer,
and $\b=(\b_0,\b_1,...,\b_m)^\top$ are the weights that connect the $z_j$ to the output which we write as 
\[\mathfrak{N}(x,w,\beta) =\b_0+\b_1z_1+...+\b_mz_m=\beta_0+\wt\beta^\top z\]
with $\wt\beta=(\beta_1,...,\beta_m)^\top$ and $z=(z_1,...,z_m)^\top$.

We assume that the conditional density $p(y|x)$ has an exponential family form
\beq\label{eq:exponential family}
p(y|x)=\exp\left(\frac{y\varpi-b(\varpi)}{\phi}+c(\phi,y)\right)
\eeq
with the canonical parameter $\varpi$ and the dispersion parameter $\phi$.
Let $g(\cdot)$ be the link function that links the conditional mean $\xi=\xi(x)=\E(y|x)$ to a function of the covariates $x$.
In the conventional GLM, $g(\xi)$ is assumed to be a linear combination of $x$.
In order to achieve flexibility and capture possible non-linear effects that $x$ has on $\xi$, we propose the more flexible model
\beq\label{eq:fGLM1}
g(\xi) = \beta_0+\wt\b^\top z=\mathfrak{N}(x,w,\beta).
\eeq
In our later examples we use the canonical link function where $\varpi=g(\xi)$, which is the log link for Poisson responses and the logit link
for binomial responses.  
The predictors (learned features) $z_j$ efficiently capture the important non-linear effects
of the original raw covariates $x$ on $g(\xi)$.

The model \eqref{eq:fGLM1} is flexible, but can be hard to interpret.  It may be useful in this respect to introduce additional structure.
One possibility is to partition the covariates as $x=({x^{(1)}}^\top,{x^{(2)}}^\top)^\top$ with $x^{(1)}$ and $x^{(2)}$ expected to have nonlinear and linear effects respectively on $g(\xi)$, 
so that 
\bea\label{eq:fGLM2}
g(\xi)&=&\mathfrak{N}(x^{(1)},w,\b^{(1)})+{\b^{(2)}}^\top x^{(2)}
\eea
where $\b^{(1)}$ parametrizes the nonlinear effects w.r.t. the covariates $x^{(1)}$ and $\b^{(2)}$ parametrizes
the linear effects w.r.t. the covariates $x^{(2)}$. Write $\beta=(\b^{(1)},\b^{(2)})$. 
We refer to the general regression model with the exponential family for response distribution 
\eqref{eq:exponential family} and the mean model \eqref{eq:fGLM1} or \eqref{eq:fGLM2} as DeepGLM.

The vector of model parameters $\theta$ consists of $w$, $\beta$ and possibly dispersion parameters $\phi$ if $\phi$ is unknown.
The density $p(y|x)$ in \eqref{eq:exponential family} is now a function of $\theta$, $p(y|x)=p(y|x,\theta)$.
Given a dataset $D$, the likelihood function is
\beq
L(\theta)=\prod_{i=1}^np(y_i|x_i,\theta), 
\eeq
and likelihood-based inference methods, including Bayesian methods, can be applied.  

\subsection{DeepGLMM}\label{sec:DeepGLMM}
Consider a panel dataset $D=\{(y_{it},x_{it}),t=1,...,T_i,i=1,...,n\}$ with $y_{it}$ the response
and $x_{it}$ the vector of covariates of subject $i$ at time $t$.
Generalized linear mixed models (GLMM) use random effects to account for within-subject dependence.
Let $z_{it,j}=\phi_j(x_{it},w)$, $j=1,...,m$, be the units in the last hidden layer of a neural net,
$z_{it}=(z_{it,1},...,z_{it,m})^\top$.
Similarly to GLMM, to account for within-subject dependence, we propose to link the conditional mean of $y_{it}$ given $x_{it}$, $\xi_{it}=\E(y_{it}|x_{it})$, to the predictors $z_{it,j}$ as follows
\bea\label{eq:mu_it}
g(\xi_{it})=\b_0+\a_{i0}+(\b_1+\a_{i1})z_{it,1}+...+(\b_m+\a_{im})z_{it,m}=\mathfrak{N}(x_{it},w,\b+\a_i),
\eea
where $\a_i=(\a_{i0},...,\a_{im})^\top$ are random effects that reflect the characteristics of subject $i$.
The variation between the subjects is captured in the distribution of $\alpha_i$. Our paper assumes $\a_i\sim \N(0,\Gamma)$ but more flexible distributional
specifications for the random effects can also be considered.

Similarly to the previous section, a more interpretable model can be developed if some additional structure is assumed.
Partition the covariates $x_{it}$ as $({x_{it}^{(1)}}^\top,{x_{it}^{(2)}}^\top)^\top$ where $x_{it}^{(1)}$ and $x_{it}^{(2)}$ are expected to have non-linear and linear effects 
respectively on $g(\xi_{it})$:  
\bea\label{eq:mu_it1}
g(\xi_{it})&=&\mathfrak{N}(x_{it}^{(1)},w,\b^{(1)})+(\a_i+\b^{(2)})^\top x_{it}^{(2)},
\eea
where $\b^{(1)}$ parametrizes fixed non-linear effects for $x_{it}^{(1)}$, and $\b^{(2)}$ and $\a_i$ are fixed and random linear effects w.r.t. the covariates $x_{it}^{(2)}$. Write $\beta=(\b^{(1)},\b^{(2)})$. The model parameters $\theta$ include $w$, $\beta$, $\Gamma$ and any dispersion parameters $\phi$ if unknown.
We refer to the panel data model with the distribution \eqref{eq:exponential family}
and the link \eqref{eq:mu_it} or \eqref{eq:mu_it1} as DeepGLMM.

The likelihood for the DeepGLMM is
\beq\label{eq:DeepGLMM likelihood 1}
L(\t)=\prod_{i=1}^nL_i(\t)
\eeq
with the $i$th likelihood contribution
\bea\label{eq:DeepGLMM likelihood 2}
L_i(\t)=p(y_i|x_i,\t)&=&\int p(y_i|x_i,w,\b,\phi,\a_i)p(\a_i|\Gamma)d\alpha_i\notag\\
&=&\int\prod_{t=1}^{T_i}p(y_{it}|z_{it},\b,\phi,\a_i)p(\a_i|\Gamma)d\alpha_i.
\eea
Section \ref{sec:SVB} describes a VB algorithm for fitting the DeepGLMM.

The likelihood for the DeepGLMM described in \eqref{eq:DeepGLMM likelihood 1} and \eqref{eq:DeepGLMM likelihood 2} is intractable,
because the integral in \eqref{eq:DeepGLMM likelihood 2} cannot be computed analytically,
except for the case where the conditional distribution of $y_{it}$ given $z_{it}$ is normal.
However, we can estimate each likelihood contribution $L_i(\theta)$ unbiasedly using importance sampling, and this allows
estimation methods for intractable likelihoods, such as the block pseudo-marginal MCMC of \cite{Tran:2016} or 
the VB method of \cite{Tran:2017}, to be used.
We consider here an alternative VB method based on the reparametrization trick in Section \ref{sec: Reparameterization trick},
which utilizes the information of the gradient of the log-likelihood computed by the back-propagation algorithm \citep{Goodfellow:2016}.
By Fisher's identity \citep{Gunawan:2017}, the gradient of the log-likelihood contribution $\nabla_\t\ell_i(\t)$,
$\ell_i(\t)=\log L_i(\t)$ with $L_i(\t)$ in \eqref{eq:DeepGLMM likelihood 2}, is
\beq\label{eq:score}
\nabla_{\theta}\ell_{i}(\theta)=\int \nabla_\t\left(\log\prod_{t=i}^{T_i}p(y_{it}|z_{it},\b,\phi,\a_i)p(\a_i|\Gamma)\right)p(\a_i|\t,y_{i},x_i)d\a_i,
\eeq
where $p(\a_i|\t,y_{i},x_i)$ is the conditional distribution of the random effects $\a_i$ given data $(y_i,x_i)$ and $\t$.
The gradient inside the integral \eqref{eq:score} can be computed by back-propagation,
and then the integral can be estimated easily by importance sampling.

\section{Gaussian variational approximation with factor covariance structure}\label{sec:SVB}
This section describes the NAtural gradient Gaussian Variational Approximation with factor Covariance method (NAGVAC) for approximate Bayesian inference in high-dimensional models.
We note that this estimation method can be used for training other high-dimensional models rather than the DeepGLM and DeepGLMM described in Section \ref{Flexible regression models with DFNN}. Let $D$ be the data and $\t\in\Theta$ the vector of unknown parameters.
Bayesian inference about $\t$ is based on the posterior distribution with density function
\[\pi(\t)=p(\t|D)=\frac{p(\t)L(\t)}{p(D)}\]
with $p(\t)$ the prior, $L(\t)=p(D|\t)$ the likelihood function and $p(D)=\int p(\theta)L(\t)d\t$ the marginal likelihood.
In all but a few simple cases the posterior $\pi(\t)$ is unknown, partly because $p(D)$ is unknown, which makes it challenging to carry out Bayesian inference.

In this work, we are interested in variational approximation methods, which are widely used as a scalable and computationally effective method
for Bayesian computation \citep{Bishop:2006,blei+kj17}.
We will approximate the posterior $\pi(\t)$ by a Gaussian distribution with density $q_\l(\t) =\N(\t;\mu,\Sigma)$, the density of a multivariate normal distribution with mean vector $\mu$ and covariance matrix $\Sigma$.
The optimal variational parameter $\l=(\mu,\Sigma)$ is chosen by minimizing the Kullback-Leibler divergence between $q_\l(\t)$ and $\pi(\t)$
\bea\label{eq:KL lambda}
\text{KL}(\l)&=&\int q_\lambda(\t)\log\frac{q_\lambda(\t)}{\pi(\t)}d\t=\int q_\lambda(\t)\log\frac{q_\lambda(\t)}{p(\t)L(\t)}d\t+\log p(D)\notag\\
&=&-\LB(\l)+\log p(D),
\eea
where
\beq\label{eq: lower bound}
\LB(\l)=\int q_\l(\t)\log\frac{p(\t)L(\t)}{q_\l(\t)}d\t
\eeq
is a lower bound on $\log\; p(D)$. Minimizing $\Kl(\l)$ is therefore equivalent to maximizing the lower bound $\LB(\l)$.
If we can obtain an unbiased estimator $\wh{\nabla_\l\LB(\l)}$ of the gradient of the lower bound, then
we can use stochastic optimization to maximize ${\LB}(\lambda)$, as in Algorithm \ref{algorithm 1} below.
\begin{algorithm}\label{algorithm 1}
\begin{itemize}
  \item Initialize $\l^{(0)}$ and stop the following iteration if the stopping criterion is met.
  \item For $t=0,1,...$, compute $\l^{(t+1)}=\l^{(t)}+a_t \wh{\nabla_\l{\LB}}(\l^{(t)})$. 
\end{itemize}
\end{algorithm}
The learning rate sequence $\{a_t\}$ in Algorithm \ref{algorithm 1} should satisfy the Robbins-Monro conditions, $a_t>0$, $\sum_t a_t=\infty$ and $\sum_t a_t^2<\infty$
\citep{Robbins:1951}. The choice of $a_t$ is discussed later on in some detail.

\subsection{Reparametrization trick}\label{sec: Reparameterization trick}
As is typical of stochastic optimization algorithms,
the performance of Algorithm \ref{algorithm 1} depends greatly on the variance of the noisy gradient
so that variance reduction methods are needed.
We will use the so-called reparametrization trick \citep{Kingma:2013,rezende+mw14} in this paper, and its modification by \cite{Roeder2017}, who
generalized ideas considered in \cite{Han2016} and \cite{Tan2016}.

Suppose that for $\t\sim q_\l(\cdot)$, there exists a deterministic function $g(\l,\eps)$ such that $\t=g(\l,\eps)\sim q_\l(\t)$ where $\eps\sim p_\eps(\cdot)$, which is independent of $\l$.
For example, if $q_\l(\t)=\N(\t;\mu,\Sigma)$ then $\t=\mu+\Sigma^{1/2}\eps$ with $\eps\sim\N(0,I)$ and $I$ is the identity matrix. 
Writing $\LB(\l)$ as an expectation with respect to $p_\epsilon(\cdot)$ gives
\begin{align*}
  \LB(\l) & = \E_\eps\Big(h(g(\eps,\l))-\log q_\l(g(\eps,\l))\Big),
\end{align*}
where $\E_\eps(\cdot)$ denotes expectation with respect to $p_\eps(\cdot)$, $h(\t):=\log(p(\t)L(\t))$.  Differentiating under the integral sign and simplifying as in \citet{Roeder2017} gives
\begin{align}
  \nabla_\l\LB(\l) & = \E_\eps\Big(\nabla_\l g(\l,\eps)\nabla_\theta\left\{h(g(\eps,\l))-\log q_\l(g(\eps,\l))\right\}\Big). \label{roedergrad}
\end{align}
The gradient (\ref{roedergrad}) can be estimated unbiasedly using  i.i.d samples $\eps_s\sim p_\eps(\cdot)$, $s=1,...,S$, as
\begin{align}
\wh{\nabla_\l{\LB}}(\l) & =\frac1S\sum_{s=1}^S\nabla_\l g(\l,\eps_s) \nabla_\t \big\{h(g(\l,\eps_s))-\log q_\l (g(\l,\eps_s))\big\}. \label{roedergradest}
\end{align}
The gradient estimator (\ref{roedergradest}) has the advantage that if the variational family is rich enough to contain the exact posterior, so that
$\exp(h(\theta))\propto q_\l(\theta)$ at the optimal $\lambda$, then the estimator (\ref{roedergradest}) is exactly zero at this optimal value even for $S=1$ 
where we use just a single
Monte Carlo sample from $p_\eps(\eps)$.  Reparametrized gradient estimators are often more efficient than alternative approaches to estimating
the lower bound gradient, partly because they take into account information from $\nabla_\theta h(\theta)$.  For further discussion we refer the reader
to \citet{Roeder2017}.

\subsection{Natural gradient}
It is well-known that the ordinary gradient $\nabla_\l{\LB}(\l)$ does not adequately capture the geometry of the approximating family $q_\l(\t)$ \citep{Amari:1998}.
A small Euclidean distance between $\l$ and $\l'$ does not necessarily mean a small \KL{} divergence between $q_\l(\t)$ and $q_{\l'}(\t)$.
\cite{Rao:1945} was the first to point out the importance of information on the geometry of the manifold of a statistical model
and introduced the Riemannian metric on this manifold induced by the Fisher information matrix. 
\cite{Amari:1998} shows that the steepest direction for optimizing the objective function $\LB (\l)$ on the manifold formed by the family $q_\l(\t)$ is directed by the so-called natural gradient 
which is defined by pre-multiplying the ordinary gradient with the inverse of the Fisher information matrix
\beq\label{eq:natural gradient}
\nabla_{\lambda}\LB (\l)^{\text{nat}} = I_F^{-1}(\l)\nabla_\l\LB(\l),
\eeq
with  $I_F(\lambda)=\cov_{q_\l}(\nabla_\l\log q_\l(\t))$.   

The use of the natural gradient in VB algorithms is considered, among others, by \cite{Sato:2001}, \cite{Honkela:2010},
\cite{Hoffman:2013}, \cite{Salimans:2013} and \cite{Tran:2017}.  
A simple demonstration of the importance of the natural gradient can be found in \cite{Tran:2017}.
The use of the natural gradient in deep learning problems is considered in \cite{Pascanu:2014}, who show
the connection between natural gradient descent and other second-order optimization methods such as Hessian-free optimization.

The main difficulty of using the natural gradient is the computation of $I_F(\lambda)$, and the solution of 
linear systems involving this matrix, which is required to compute \eqref{eq:natural gradient}. 
The problem is more severe in high dimensional models because this matrix often has a large size.
Some approximation methods, such as the truncated Newton approach, are needed \citep{Pascanu:2014}.
We consider in the next section an efficient method for computing $I_F(\lambda)^{-1}\nabla_\l \LB(\l)$ based on the use of iterative conjugate gradient methods
for solving linear systems when the covariance matrix of the Gaussian variational approximation is parametrized by a factor model.  We compute \eqref{eq:natural gradient} 
by solving the linear system $I_F(\lambda)x=\nabla_\lambda \LB (\l)$ for $x$ using only matrix-vector products involving $I_F(\lambda)$, where the matrix
vector products can be done efficiently both in terms of computation time and memory requirements by using the factor structure
of the variational covariance matrix.
In the special cases of one factor the natural gradient in \eqref{eq:natural gradient} can be computed analytically and efficiently.

\subsection{Gaussian variational approximation with factor covariance}\label{sec:Gaussian VB with factor decomposition}
We now describe in detail the Gaussian variational approximation with factor covariance (VAFC) method of \citet{ong+ns16}.  The VAFC method considers the multivariate
normal variational family $q_\lambda(\theta)=\N(\mu,\Sigma)$, where $\Sigma$ is parametrized as 
\beq\label{e:factor decomposition}
\Sigma=BB^\top+D^2.
\eeq
The factor loading matrix $B$ is of size $d\times f$,
where $d$ is the dimension of $\theta$ and $f$ the number of factors, $f\ll d$, and $D$ is diagonal with diagonal entries $c=(c_1,\dots, c_d)^\top$.  $c$ is 
a vector of idiosyncratic noise standard deviations.  Factor structures
are well known to provide useful parsimonious representations of dependence in high-dimensional settings \citep{Bartholomew2011}.  We assume $B$ is lower triangular, i.e.,
$B_{ij}=0$ for $j>i$.  Although imposing the constraint $B_{ii}>0$ makes the factor representation identifiable \citep{Geweke1996}, we do not impose
this constraint to simplify the optimization. The variational optimization simply locks onto one of the equivalent modes.  
An intuitive generative representation of the factor structure that is the basis of our application of the reparametrization trick is the following:  
if we consider $\theta\sim q_\l(\t)=\N(\mu,BB^\top+D^2)$, then we can represent $\theta$ as 
$\theta=\mu+B\epsilon_1+c\circ \epsilon_2$ where $\epsilon=(\epsilon_1^\top,\epsilon_2^\top)^\top\sim \N(0,I)$, $\epsilon_1$ and $\epsilon_2$ have dimensions $f$ and $d$ respectively, and $\circ$ denotes the Hadamard (element by element) product for two matrices of the same size.
We can see from this representation that the latent variables $\epsilon_1$ (the ``factors", which are low-dimensional) explain all the correlation between
the components, whereas component-specific idiosyncratic variance is being captured through $\epsilon_2$.  

The variational parameters are $\l=(\mu^\top,\vec (B)^\top,c^\top)^\top$, where we have written $\vec (B)$ for the vectorization of $B$ obtained by stacking
its columns from left to right.  \citet{ong+ns16} show that the gradient of the lower bound takes the form
\begin{align}
 \nabla_\mu \LB(\l)= & \E_\epsilon\Big(\nabla_\theta h(\mu+B\epsilon_1+c\circ\epsilon_2)+(BB^\top+D^2)^{-1}(B\epsilon_1+c\circ\epsilon_2)\Big), \label{gradmu}\\
  \nabla_B \LB(\l) = & \E_\epsilon\Big(\nabla_\theta  h(\mu+B\epsilon_1+c \circ\epsilon_2)\epsilon_1^\top+(BB^\top+D^2)^{-1}(B\epsilon_1+c\circ\epsilon_2)\epsilon_1^\top\Big),  \label{gradB}
\end{align}
and
\begin{align}
  \nabla_c \LB(\l) = & \E_\epsilon\Big(\diag\big(\nabla_\theta h(\mu+B\epsilon_1+c\circ\epsilon_2)\epsilon_2^\top+(BB^\top+D^2)^{-1}(B\epsilon_1+c\circ\epsilon_2)\epsilon_2^\top\big)\Big),  \label{gradd}
\end{align}
where we have written $\diag (A)$ for the diagonal elements of a square matrix $A$. 
Here, the inverse matrix $(BB^\top+D^2)^{-1}$ can be computed efficiently; see \eqref{eq:Woodbury}.
We note that in the expression for the gradient of $B$ above, we should set to zero the upper triangular components which correspond to elements of $B$ which
are fixed at zero.  Unbiased estimation of gradients for stochastic gradient ascent can proceed based on these expressions by drawing one or more samples from
$p_\epsilon(\cdot)$ to estimate the expectations.  

\subsection{Efficient natural gradient VAFC method}\label{sec:Efficient natural gradient}
We now describe how to efficiently compute the natural gradient \eqref{eq:natural gradient}
by leveraging the factor structure \eqref{e:factor decomposition}.
\citet{ong+ntsd17} also
considered a natural gradient method for Gaussian variational approximation with factor covariance.  However, this was in the context
of likelihood-free inference methods where the dimension of $\theta$ is low compared to the models of interest here, and they simply used naive methods for solving the
linear systems involving $I_F(\lambda)$ required to compute the natural gradient.
This is impractical in high-dimensional problems, and here we demonstrate how to implement natural
gradient VAFC when $\theta$ is high-dimensional using conjugate gradient methods (see, for example, \citet{Stoer1983}).  

Write $I_F(\lambda)$ in partitioned form as 
$$I_F(\lambda)=\left[ \begin{array}{ccc} I_{11} & I_{21}^\top & I_{31}^\top \\ I_{21} & I_{22} & I_{32}^\top \\
I_{31} & I_{32} & I_{33} \end{array} \right],$$
where the blocks in the partition follow the partition of $\lambda$ as $\lambda=(\mu^\top,\vec (B)^\top,c^\top)^\top$. 
Because the upper triangle of $B$ is fixed at zero the corresponding
rows and columns of $I_F(\lambda)$ should be omitted.  \citet{ong+ntsd17} show that
$I_{11}=\Sigma^{-1}$, $I_{21}=I_{31}=0$, $I_{22}=2(B^\top\Sigma^{-1}B\otimes \Sigma^{-1})$ (where $\otimes$ denotes the Kronecker product), 
$I_{33}=2 (D\Sigma^{-1})\circ (\Sigma^{-1}D)$ and $I_{32}=2(B^\top\Sigma^{-1}D\otimes \Sigma^{-1})E_d^\top$, where $E_d$ is the $d\times d^2$ matrix
that picks out the diagonal elements of the $d\times d$ matrix $A$ from its vectorization, so that $E_d\vec (A)=\diag (A)$.  
To use a conjugate gradient linear solver to compute $I_F(\l)^{-1}\nabla_\lambda \LB(\l)$ we simply need to be able to compute matrix vector
products of the form $I_F(\lambda) x$ for any vector $x$ quickly without needing to store the elements of $I_F(\lambda)$.  

This can be done provided we can do matrix vector products for the matrices $I_{11}$, $I_{22}$, $I_{33}$ and $I_{32}$. 
Except for the one-factor case described below, this is still difficult so we further approximate
$I_F(\lambda)$ by setting $I_{32}=0$ and replacing $I_{33}$ with
$\tilde{I}_{33}=2(D \tilde{\Sigma}^{-1})\circ (\tilde{\Sigma}^{-1}D)$, where $\tilde{\Sigma}$ is the diagonal approximation to $\Sigma$ obtained by setting
its off-diagonal elements to zero.  Using this approximation and $I_{32}=0$ we obtain a positive definite approximation $\tilde{I}_F(\lambda)$ to $I_F(\lambda)$ which
we use instead of $I_F(\lambda)$ in the natural gradient.  
We note that these approximations do not affect the factor structure of $\Sigma$ in \eqref{e:factor decomposition}.

Multiplications involving $\tilde{I}_{33}$ are simple since this matrix is diagonal, but we still need efficient methods to compute
matrix vector products for $I_{11}$ and $I_{22}$.  Considering $I_{11}$ first, we note that by using the Woodbury formula 
\beq\label{eq:Woodbury}
  I_{11}=\Sigma^{-1}  = D^{-2}-D^{-2}B(I+B^\top D^{-1} B)^{-1}B^\top D^{-2},
\eeq
and then noting that $D$ is diagonal and $(I+B^\top D^{-2}B)$ is $f\times f$, $f\ll d$, we can calculate $I_{11}x=\Sigma^{-1}x$ without needing to store
any $d\times d$ matrix or do any dense $d\times d$ matrix multiplications.  Next, consider $I_{22}x$ for some vector $x$.  We note that
\begin{align*}
  I_{22}=2(B^\top\Sigma^{-1}B\otimes \Sigma^{-1})x & = 2\vec (\Sigma^{-1} x^* B^\top\Sigma^{-1}B),
\end{align*}
where $x^*$ denotes the $d\times f$ matrix such that $x=\vec (x^*)$ and where we have used the identity that for conformable matrices
$X,Y,Z$, $\vec (XYZ)=(Z^\top\otimes X)\vec (Y)$.  Then $\Sigma^{-1}x^*$ is computed efficiently by the Woodbury formula, and similarly for 
$B^\top\Sigma^{-1}B$.  
We refer to our natural gradient estimation method with $f$ factors as NAGVAC-$f$.

We now consider the special case of the NAGVAC-$1$ method where the covariance matrix $\Sigma$ is parameterized as in \eqref{e:factor decomposition} with $B$ a vector.
In this case, the natural gradient \eqref{eq:natural gradient} can be computed efficiently
and the computational complexity is $O(d)$. See Algorithm \ref{algorithm 2}, whose detailed derivation is in the Appendix.
This estimation method is computationally attractive, especially when the dimension $d$ is extremely large.
The experimental studies in Section \ref{Applications} suggest that in some applications this method is able to produce a prediction accuracy 
comparable to the accuracy obtained by methods that use more flexible factor decomposition structures of $\Sigma$.
\cite{Trippe18} discuss the phenomenon where richer variational families produce inferior performance in terms of predictive loss for
neural networks models, and provide some theoretical insights into this phenomenon.

\begin{algorithm}[Computing the natural gradient]\label{algorithm 2}
Input: Vector $B$, $c$ and ordinary gradient vector  $g = (g_1^\top,g_2^\top,g_3^\top)^\top$ with $g_1$ the vector formed by the first $d$ elements of $g$,
$g_2$ formed by the next $d$ elements, and $g_3$ the last $d$ elements. 
Output: The natural gradient $g^{\text{nat}}=I_F^{-1}g$.
\begin{itemize}
\item Compute the vectors $v_1=c^2-2B^2\circ c^{-4}$, $v_2=B^2\circ c^{-3}$, and the scalars $\kappa_1=\sum_{i=1}^db_i^2/c_i^2$, $\kappa_2=\frac{1}{2}(1+\sum_{i=1}^d v_{2i}^2/v_{1i})^{-1}$.
\item Compute 
\[g^{\text{nat}}=\begin{pmatrix}
(g_1^\top B)B+c^2\circ g_1\\
\frac{1+\kappa_1}{2\kappa_1}\Big((g_2^\top B)B+c^2\circ g_2\Big)\\
\frac12v_1^{-1}\circ g_3+\kappa_2 \big[(v_1^{-1}\circ v_2)^\top g_3\big](v_1^{-1}\circ v_2)
\end{pmatrix}.\]
\end{itemize}
\end{algorithm}

\section{Variable selection and regularization priors}\label{Variable selection}
This section first presents a method for variable selection in DeepGLM and DeepGLMM.
The method is based on the Bayesian adaptive group Lasso method which is developed for GLM in \cite{Leng:2013} and normal regression in \cite{Kyung:etal:2010}.
Consider a neural network as in Figure \ref{f:neural net}.
Denote by $w_{X_j}$ the vector of weights that connect the covariate $X_j$ to the $m$, say, units in the first hidden layer. 
We use the following priors 
\beqn
w_{X_j}|\tau_j\sim \N(0,\tau_jI_m),\;\;\;\;\tau_j|\gamma_{j}\sim\text{Gamma}\left(\frac{m+1}{2},\frac{\gamma_{j}^2}{2}\right),\;\;j=1,...,p,
\eeqn
with the $\gamma_j>0$ the shrinkage parameters.
By the normal-Gamma mixture \citep{Andrews:1974,Kyung:etal:2010}, we have that
\[p(w_{X_j}|\gamma_j)=\int p(w_{X_j}|\tau_j)p(\tau_j|\gamma_{j})d\tau_j\propto\exp\big(-\gamma_{j}\|w_{X_j}\|_{2}\big)\]
with $\|w_{X_j}\|_{2}$ the $l_2$-norm of $w_{X_j}$. Hence, the posterior mode of the $w_{X_j}$ induced from the above hierarchical prior with the same $\gamma_{j}=\gamma$ is equivalent to the group Lasso estimator of \cite{Yuan:2006}.

Selecting the adaptive shrinkage parameters $\gamma_j$ is challenging. We develop an empirical Bayes method
for estimating these tuning parameters based on an iterative scheme within the variational approximation procedure.
Writing the Bayesian hierarchical model in the generic form
\beqn
y|\psi,\theta\sim p(y|\t,\psi),\;\;\;\t|\psi\sim p(\t|\psi),
\eeqn 
with $y$ the data, $\t$ the model parameters and $\psi$ the hyperparameters to be selected.
The marginal likelihood for $\psi$ is $p(y|\psi)=\int p(\t|\psi)p(y|\t,\psi)d\t$, which can be maximized using an EM-type algorithm \citep{Casella:2001}.
Given an initial value $\psi^{(0)}$, we iteratively update $\psi$ by
\[\psi^{(k+1)}=\argmax_\psi\left\{\E_{\t|y,\psi^{(k)}}\log p(y,\t|\psi)\right\},\]
where $\E_{\t|y,\psi^{(k)}}(\cdot)$ is the expectation with respect to the posterior distribution $p(\t|y,\psi^{(k)})$.
It can be shown that the updating rule for $\gamma_j$ is
\beq\label{eq:update gamma_j}
\gamma_j^{(k+1)}=\sqrt{\frac{m+1}{\E_{\t|y,\gamma_j^{(k)}}[\tau_j]}}.
\eeq
In our variational approximation framework, the expectation $\E_{\t|y,\psi^{(k)}}(\cdot)$ can be naturally approximated by the expectation w.r.t. the current variational approximation $q_{\l^{(k)}}(\theta)$,
and the updates in \eqref{eq:update gamma_j} can be computed in closed form.
One can merge the auxiliary parameters $\tau_1,...,\tau_p$ into the model parameters and learn them jointly by the Gaussian variational posterior $q_{\l}(\theta)$.
Alternatively, one can conveniently update the variational posterior for each $\tau_j$ separately in a fixed-form within mean-field variational approximation procedure; see, e.g. \cite{Tran:2016JCGS}.
We use the latter in this paper. The optimal VB posterior for $1/\tau_j$ is inverse-Gaussian with mean and shape parameter
\beq\label{eq:update alpha beta_j}
\alpha_{\tau_j}\leftarrow\frac{\gamma_j}{\sqrt{\E_{q_\lambda}[w_{X_j}'w_{X_j}]}},\;\;\;\beta_{\tau_j}\leftarrow\gamma_j^2.
\eeq
Because $\E_{\t|y,\gamma_j^{(k)}}[\tau_j]$ can be approximated by
$\E_{q_\lambda}[\tau_j]=1/\alpha_{\tau_j}+1/\beta_{\tau_j}$, the update in \eqref{eq:update gamma_j} becomes   
\beq\label{eq:update gamma_j new}
\gamma_j\leftarrow\sqrt{\frac{m+1}{1/\alpha_{\tau_j}+1/\beta_{\tau_j}}}.
\eeq
The updates in \eqref{eq:update alpha beta_j} and \eqref{eq:update gamma_j new} are then embedded in the main variational iterate procedure Algorithm 1.
This leads to a convenient and principled way for selecting the shrinkage parameters $\gamma_j$,
which would be very challenging for alternative methods.

In order to control overfitting, we use a ridge-type regularization for the rest of the weights in the neural network.
Let $\wt w$ be the vector of all the weights except those that connect the input layer to the first hidden layer and the intercepts (also called bias terms).
We use the following prior for $\wt w$
\[p(\wt w)\propto\exp\left(-\frac{\gamma_w}{2}\wt w^\top \wt w\right),\]
with $\gamma_w$ the shrinkage parameter.
Similarly to above, $\gamma_w$ is selected by maximizing the marginal likelihood and the update rule is
\beq\label{eq:EM_like gamma w}
\gamma_w\leftarrow\frac{d_{\wt w}}{\E_{q_\lambda}[\wt w^\top \wt w]},
\eeq
where $d_{\wt w}$ denotes the dimension of $\wt w$.

\section{Practical recommendations in training DeepGLM and DeepGLMM}\label{Bayesian treament}
The NAGVAC estimation method can be used as a general estimation method for any model.
However, this section focuses on estimating DFNN-based models,
and discusses some implementation recommendations that we found useful in practice.

\subsection{Stopping rule and lower bound for model choice} 
It is common in deep learning applications of neural network methods to implement early stopping in the optimization to 
avoid overfitting, because often the training loss decreases steadily over the optimization updates, but the validation loss starts increasing at some point.
In our VB framework, the lower bound \eqref{eq: lower bound} can be monitored to check the convergence and decide when to stop training.
There are two advantages to using the lower bound.
The first advantage is that a validation set is unnecessary,
which leaves more information for the training phrase.
Also, a stopping rule based on a validation set might depend on how the validation set is selected.
The second advantage is that, given that appropriate regularization priors on the weights as the ones in Section \ref{Variable selection} have been used to control overfitting,
 the maximized lower bound can be used for model choice; see the examples in Section \ref{Applications}.
The lower bound \eqref{eq: lower bound} can be estimated efficiently using the same sample of $\theta$ generated for computing the gradient vector.
To reduce the noise level in estimating the lower bound, we follow \cite{Tran:2017} and take the average of the lower bound over a moving window of $K$ iterations.
We stop training if this moving averaged lower bound does not improve after $P$ iterations.

\subsection{Learning rate and the momentum method}
We employ the following fixed learning rate which is widely used in the deep learning literature
\beq\label{e:learningrate}
a_t=\eps_0\frac{\tau}{t},
\eeq
where $\eps_0$ is a small value, e.g. 0.01 and $\tau$ is some threshold from which the learning rate is reduced, e.g. $\tau=1000$.  
It might also be useful to use some adaptive learning rate methods, and later we consider one method adapted from \cite{Ranganath:2013}
which is described in \cite{ong+ntsd18}.  This learning rate is suitable for use with the natural gradient.  

As a method of accelerating the stochastic gradient optimization we also consider using the 
momentum method \citep{Polyak:1964}.
The update rule is 
\bean
\overline{{\nabla_\l{\LB}}} &=& \alpha_\text{m} \overline{{\nabla_\l{\LB}}}+(1-\alpha_\text{m})\wh{\nabla_{\lambda}\Kl}(\l^{(t)})^{\text{nat}},\\
\l^{(t+1)}&=&\l^{(t)}+a_t \overline{{\nabla_\l{\LB}}},
\eean
where $\alpha_\text{m}\in[0,1]$ is the momentum weight.
The use of the moving average gradient $\overline{{\nabla_\l{\LB}}}$ helps remove some of the noise
inherent in the estimated gradients of the lower bound.
See \cite{Goodfellow:2016}, Chapter 8, for a detailed discussion on the usefulness of the momentum method.

\subsection{Activation function and initialization}
For initialization of the variational parameters $\l^{(0)}=(\mu^{(0)},B^{(0)},c^{(0)})$,
we follow \cite{Glorot:2010} and initialize each weight in $\mu^{(0)}$ by the uniform distribution $\mathcal U(-\sqrt{\frac{6}{m+n}},\sqrt{\frac{6}{m+n}})$,
where the weight connects a layer with $m$ units to a layer with $n$ units.
The elements in $B^{(0)}$ are initialized by $\N(0,0.01^2)$ and  
the elements in $c^{(0)}$ are initialized by 0.01.
It is advisable to first standardize the input data so that each column has a zero sample mean and a standard deviation of one.
Finally, we use the rectified activation function $h(x)=\max(0,x)$ in all examples, unless otherwise stated. This activation function has a strong connection with biological neuroscience
and has been widely used in the deep learning literature; see, e.g., \cite{Goodfellow:2016}.

\section{Experimental studies and applications}\label{Applications}
To illustrate the performance of variable selection, the prediction accuracy of DeepGLM and DeepGLMM, and the efficiency of the NAGVAC training algorithm, we consider a range of experimental studies and applications.
All the examples are implemented in Matlab and run on a desktop computer with i5 3.3 Ghz Intel Quad Core.
All the DeepGLM and DeepGLMM models are trained using the NAGVAC-1 method, unless otherwise stated.

We use two predictive measurements. The first is the partial predictive score (PPS)
\[PPS = -\frac{1}{n_\text{test}}\sum_{(x_i,y_i)\in\text{test data}} \log p(y_i|x_i,\wh\theta),\]
with $\wh\theta$ a point estimate of the model parameters.
The second is the mean squared error (MSE),
\[MSE = \frac{1}{n_\text{test}}\sum_{(x_i,y_i)\in\text{test data}} (y_i-\wh y_i)^2,\]
with $\wh y_i$ a prediction of $y_i$, which is called the misclassification rate (MCR) for binary response $y_i$.

\subsection{Experimental studies}
\subsubsection{Efficiency of the NAGVAC algorithm}
We first study the efficiency of the NAGVAC algorithm as a genenal training method.
We use the German credit dataset from the UCI Machine Learning Repository, \texttt{http://archive.ics.uci.edu/ml}, with 1000 observations,
of which 750 are used as the training data and the rest as the test data.
The data consist of a binary response variable, credit status, which is good credit (1) or bad credit (0), together with 30 covariate variables such as education, credit amount, employment status, etc. We consider a simple logistic regression model for predicting the credit status, based on the covariates. We use an improper flat prior for the coefficients $\theta$, i.e. $p(\theta)\propto 1$. 

\begin{figure}[ht]
\centering
\includegraphics[width=165mm,height=100mm]{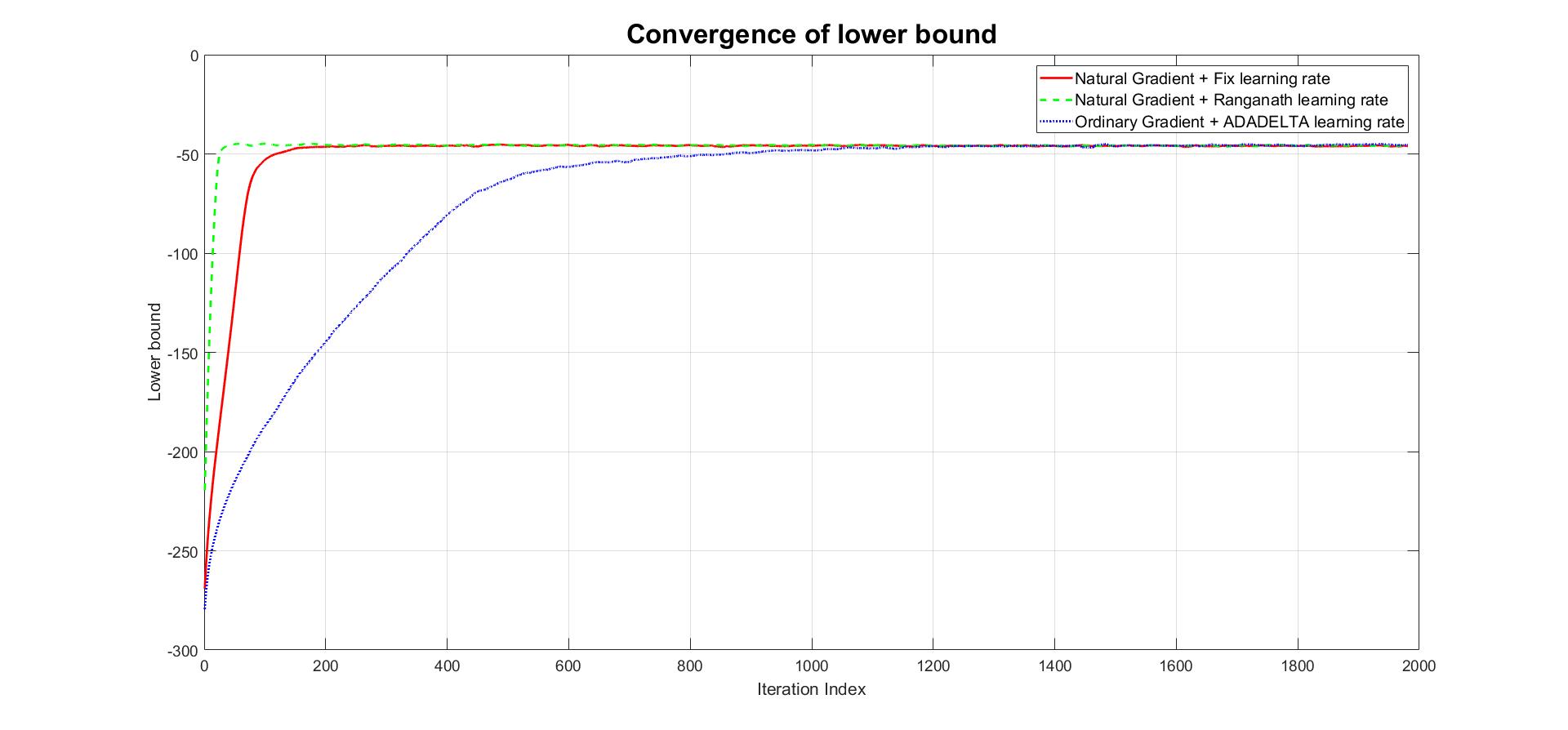}
\caption{Plot of the lower bound over iterations for the ordinary gradient and NAGVAC-1 methods.}
\label{f:GermanData2}
\end{figure}

We study the performance of the natural gradient method compared to the ordinary gradient method.
Figure \ref{f:GermanData2} shows the convergence of the lower bound of the ordinary gradient method and the NAGVAC-1 method.
For the ordinary gradient, we use the adaptive learning rate method ADADELTA of \cite{Zeiler:2012}. 
For the natural gradient, we use both the fixed learning rate in \eqref{e:learningrate}
and the adaptive learning rate of \cite{ong+ntsd18} (which is based on the method in \cite{Ranganath:2013}).
The figure shows that both the natural gradient method speed up the convergence significantly.
Although incorporating an adaptive learning rate into NAGVAC helps to speed up the convergence,
the improvement is not always significant compared to a fixed learning rate.
We used the fixed learning rate \eqref{e:learningrate} in all the examples reported below.

\subsubsection{Variable selection and prediction accuracy of DeepGLM: Binary response}
\label{sec:Prediction_accuracy_of_DL_models:binary}
Data are generated from the following deterministic model
\bean
a &=& 5 - 2(x_1+2x_2)^2 + 4x_3x_4 + 3x_5+\sum_{i=6}^{20}0x_i,\\
y &=& \begin{cases}
1,& a\geq0,\\
0,&a<0,
\end{cases}
\eean
where the $x_i$ are generated from the uniform distribution $\mathcal U(-1,1)$.
Note that the last 15 variables are irrelevant variables.
The training data consist of $n_\text{train}=100,000$ observations and the test data of $n_\text{test}=100,000$ observations. 

\begin{figure}[ht]
\centering
\includegraphics[width=165mm,height=100mm]{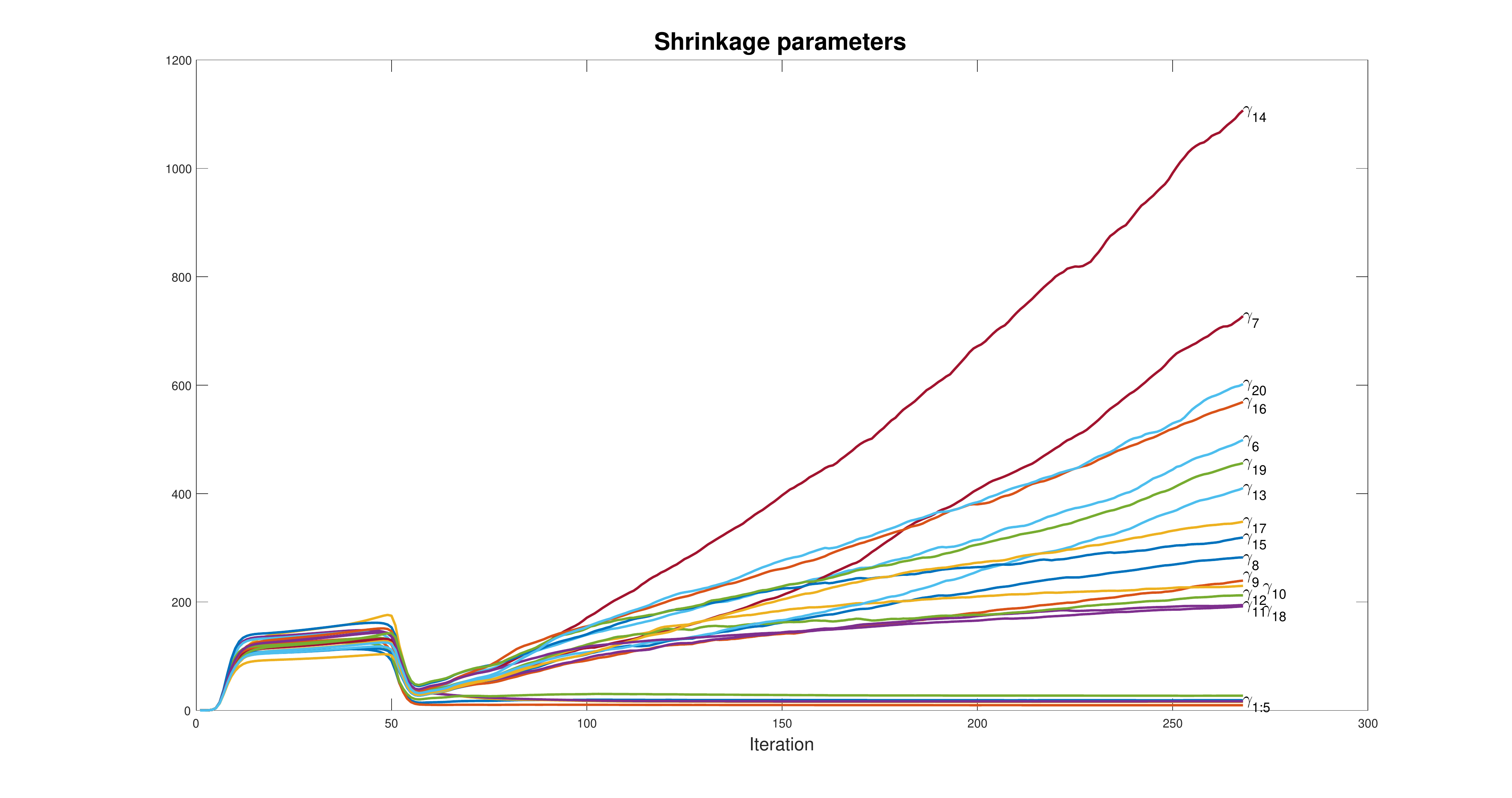}
\caption{Plots of the shrinkage parameters $\gamma_j$ over iterations. The shrinkage parameters w.r.t the irrelevant variables keep increasing, while the ones w.r.t the relevant variables keep decreasing.}
\label{f:Binary ex: shrinkage}
\end{figure}

We use a neural net with the structure (20,20,20,1),
i.e. the input layer has 20 variables, two hidden layers each has 20 units and one scalar output. 
Figure \ref{f:Binary ex: shrinkage} plots the update of the shrinkage parameters $\gamma_j$ as in \eqref{eq:update gamma_j new}.
The shrinkage parameters $\gamma_{6},...,\gamma_{20}$ w.r.t. the irrelevant variables keep increasing over iterations, while the ones w.r.t. the relevant variables keep decreasing.
This shows that the Bayesian adaptive group Lasso with the empirical Bayes method for updating the shrinkage parameters is able to scan out correctly irrelevant covariates.

We now compare the predictive performance of DeepGLM to the Bayesian Additive Regression Tree method (BART) of \cite{chipman2010}.
BART is a commonly used nonparametric regression method in the statistics literature and is well known for its prediction accuracy and its ability to capture nonlinearity effects.
Table \ref{tab: binary simulation results} summarizes the prediction performance of DeepGLM compared to that of the conventional GLM and BART. 
GLM, which is logistic regression in this example, does just slightly better than a random guess with a MCR of 40.84\%.  
As shown, DeepGLM works very well in this example and outperforms both GLM and BART.
All the comparisons with BART in this paper are done using the R package \texttt{BART} (the lattest version 1.6) with default settings for the tuning parameters.   
 
\begin{table}[h]
\begin{center}
\begin{tabular}{l|cc}
\hline\hline
Method&PPS&MCR (\%)\\
\hline
GLM&0.67&40.84\\
BART&0.08&3.09\\
DeepGLM&0.03&1.03\\
\hline\hline
\end{tabular}
\end{center}
\caption{Binary response simulation: Performance of DeepGLM v.s. GLM and BART in term of the partial predictive score (PPS) and the misclassification rate (MCR). Both are evaluated on the test data. }\label{tab: binary simulation results}
\end{table}

\subsubsection{Variable selection and prediction accuracy of DeepGLM: Continuous response}
\label{sec:Prediction_accuracy_of_DL_models}
We generate data from the following highly nonlinear model
\beq\label{eq:continuous model}
y=5 + 10x_1 + \frac{10}{x_2^2+1} + 5x_3x_4 + 2x_4 + 5x_4^2 + 5x_5 + 2x_6 + \frac{10}{x_7^2+1} + 5x_8x_9 + 5x_9^2 + 5x_{10} +\epsilon,
\eeq
where $\epsilon\sim\N(0,1)$, $(x_1,...,x_{20})^\top$ are generated from a multivariate normal distribution with mean zero and covariance matrix $(0.5^{|i-j|})_{i,j}$.
Note that the last 10 variables are irrelevant variables. The training data has 100,000 observations and the test data has 20,000.

\begin{table}[h]
\begin{center}
\begin{tabular}{lcc}
\hline\hline
Method&PPS&MSE\\
\hline
GLM&$3.31$&275.5\\
BART&1.49&4.89\\
DeepGLM&$1.20$&4.07\\
\hline\hline
\end{tabular}
\end{center}
\caption{Continuous response simulation: Performance of DeepGLM v.s. conventional GLM and BART in term of the partial predictive score (PPS) and the mean squared error (MSE). Both are evaluated on the test data. The structure of the neural net is (20,20,20,1).}\label{tab: continuous simulation results}
\end{table}

\begin{figure}[ht]
\centering
\includegraphics[width=165mm,height=100mm]{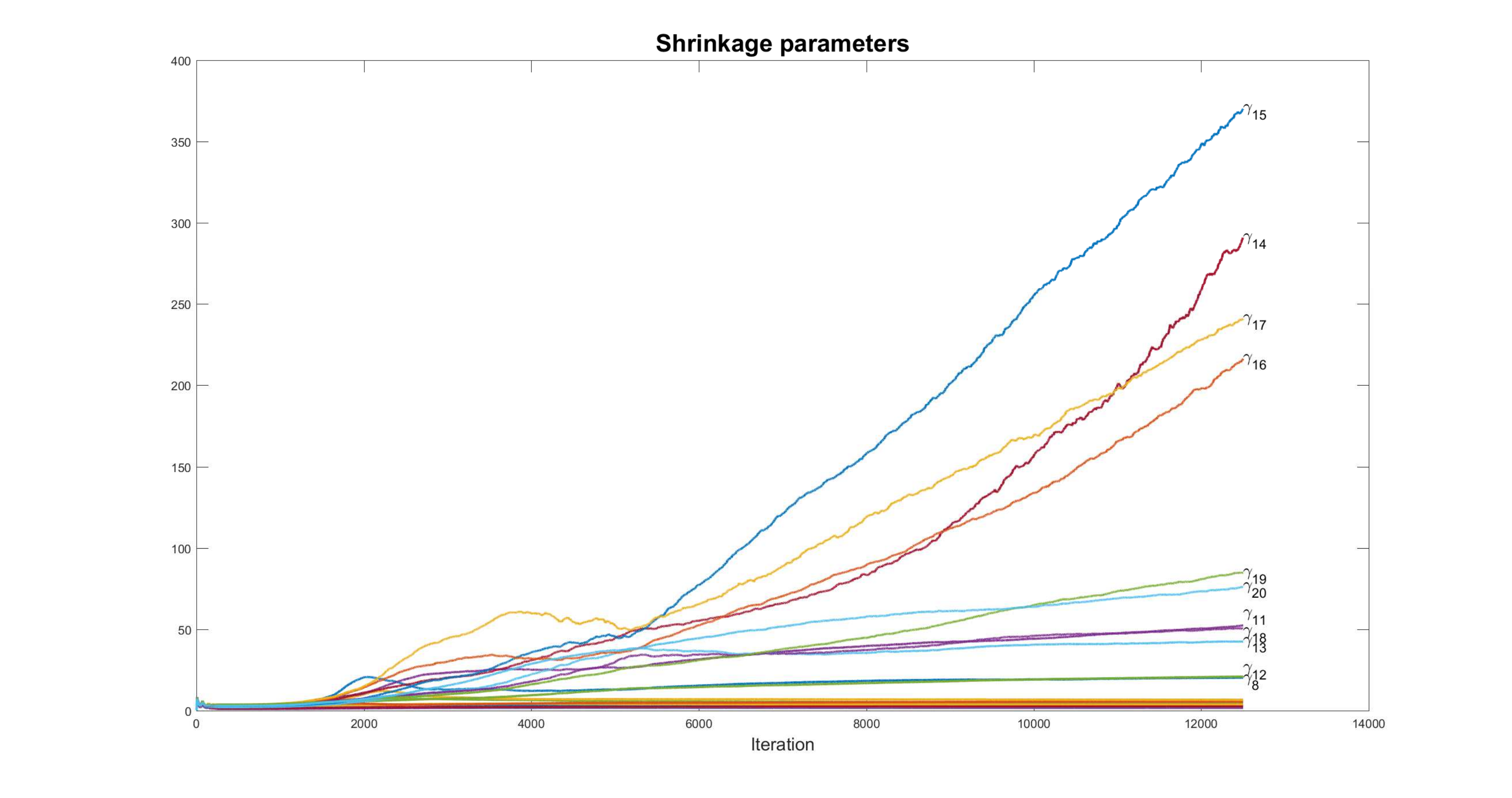}
\caption{Plots of the shrinkage parameters $\gamma_j$ over iterations. The shrinkage parameters w.r.t. the irrelevant variables keep increasing, while the ones w.r.t. the relevant variables keep decreasing.}
\label{f:continuous ex: shrinkage}
\end{figure}

Figure \ref{f:continuous ex: shrinkage} plots the update of the shrinkage parameters $\gamma_j$ as in \eqref{eq:update gamma_j new}.
All the shrinkage parameters w.r.t. the irrelevant variables keep increasing over iterations, while the ones w.r.t. the relevant variables, except $\gamma_8$, keep decreasing.
The reason is that the main effect of $x_8$ is not included into model \eqref{eq:continuous model}, hence the signal of $x_8$ might not be strong enough to be identifiable.
Table \ref{tab: continuous simulation results} summarizes the prediction performance of DeepGLM in comparison to the conventional GLM and BART. 
DeepGLM works well in this example and ourperforms both GLM and BART.

\subsubsection{Prediction accuracy of DeepGLMM: binary panel data simulation}
\label{sec:Binary_panel_data_simulation}
We study the DeepGLMM on a simulation binary panel dataset $D=\{(x_{it},y_{it});\ t=1,...,20; \ i=1,...,1000\}$ 
with $x_{it}$ the vector of covariates and $y_{it}$ the response of subject $i$ at time $t$. The response $y_{it}$ is generated from the following model:
\bean
\label{equation:binary_panel}
a_{it}&=&2+3(x_{it,1}-2x_{it,2})^2-5\frac{x_{it,3}}{(1+x_{it,4})^2}-5x_{it,5}+b_i+\eps_{it},\\
y_{it}&=&\begin{cases}
						1, & \text{if $a_{it}>0$},\\
						0, & \text{otherwise,}
\end{cases} 
\eean
where $b_i \sim \N(0,0.1)$ is a random intercept representing charactersistics of subject $i$ and $\eps_{it}\sim \N(0,1)$ is random noise associated with reponses $y_{it}$. The $x_{it,j},j=1,...,5$, are generated from a uniform distribution $\mathcal U(-1,1)$.

We fit the following DeepGLMM to this dataset
\beqn
y_{it}|x_{it}\sim\text{Binomial}(1,\mu_{it}),\;\;\;\;\log\left(\frac{\mu_{it}}{1-\mu_{it}}\right)=\mathfrak{N}(x_{it},w,\beta+\alpha_i),
\eeqn
where $\mathfrak{N}(x_{it},w,\beta+\a_i)$ is the scalar output of a neural net with input $x_{it}$, inner weights $w$ and output weights $\beta+\a_i$.
We assume that the random effects $\a_i\sim \N(0,\Gamma)$ with $\Gamma=\diag(\Gamma_0,...,\Gamma_m)$.
The model parameters are $\t=(w,\beta,\Gamma_0,...,\Gamma_m)$.
We use Gamma priors on the $\Gamma_j$, $\Gamma_j\sim\text{Gamma}(a_0,b_0)$,
and set the hyperparameters $a_0=1$ and $b_0=0.1$ in this example.
The Appendix gives further details on training this model.

For each subject, we use the first 17 observations for training and the last 3 observations for testing. That is, we are interested in the within-subject prediction. 
The neural network has one hidden layer with 10 nodes, and we compare the DeepGLMM model with the conventional logistic regression model with a random intercept.  

For binary panel data, the misclassification rate is defined as follows.
Let $\bar{\t}$ be the mean of the VB approximation $q_\l(\t)$ after convergence.
Let $\wh\mu_{\a_i}$ in \eqref{eq:mu_alpha_i} be the mode of $p(\a_i|y_i,x_i,\bar{\theta})$,
which can be used as the prediction of the random effects $\a_i$ of subject $i$.
Let $(y_{it_0},x_{it_0})$ be a future data pair.
We set the prediction of $y_{it_0}$ as
\[\wh y_{it_0}=1\text{  if and only if } \mathfrak{N}(x_{it_0},w,\beta+\wh\mu_{\a_i})\geq 0.\]
The classification error is
\[\text{MCR}=\sum I_{\wh y_{it_0}\not= y_{it_0}}/\text{total number of future observations}\]
with the sum over the test data points $(y_{it_0},x_{it_0})$. 

\begin{table}[ht]
	\begin{center}
		\begin{tabular}{lll}
			\hline\hline
			Model&PPS&MCR\\
			\hline
			GLMM&1.24&17.57\%\\
			DeepGLMM&0.13&5.27\%\\
			\hline\hline
		\end{tabular}
	\end{center}\caption{Simulation binary panel dataset: Performance of DeepGLMM v.s. GLMM in term of the partial predictive score (PPS) and the misclassification rate (MCR). Both are evaluated on the test data.}\label{tab:simulation panel binary performance}
\end{table}
Table \ref{tab:simulation panel binary performance} summarizes the performance of DeepGLMM and GLMM. The results show that modelling covariate effects in a flexible way using the neural network basis functions is helpful here in terms of improving both PPS and MCR.

\subsection{Applications}
\subsubsection{Direct Marketing data}
\begin{figure}[ht]
\centering
\includegraphics[width=165mm,height=80mm]{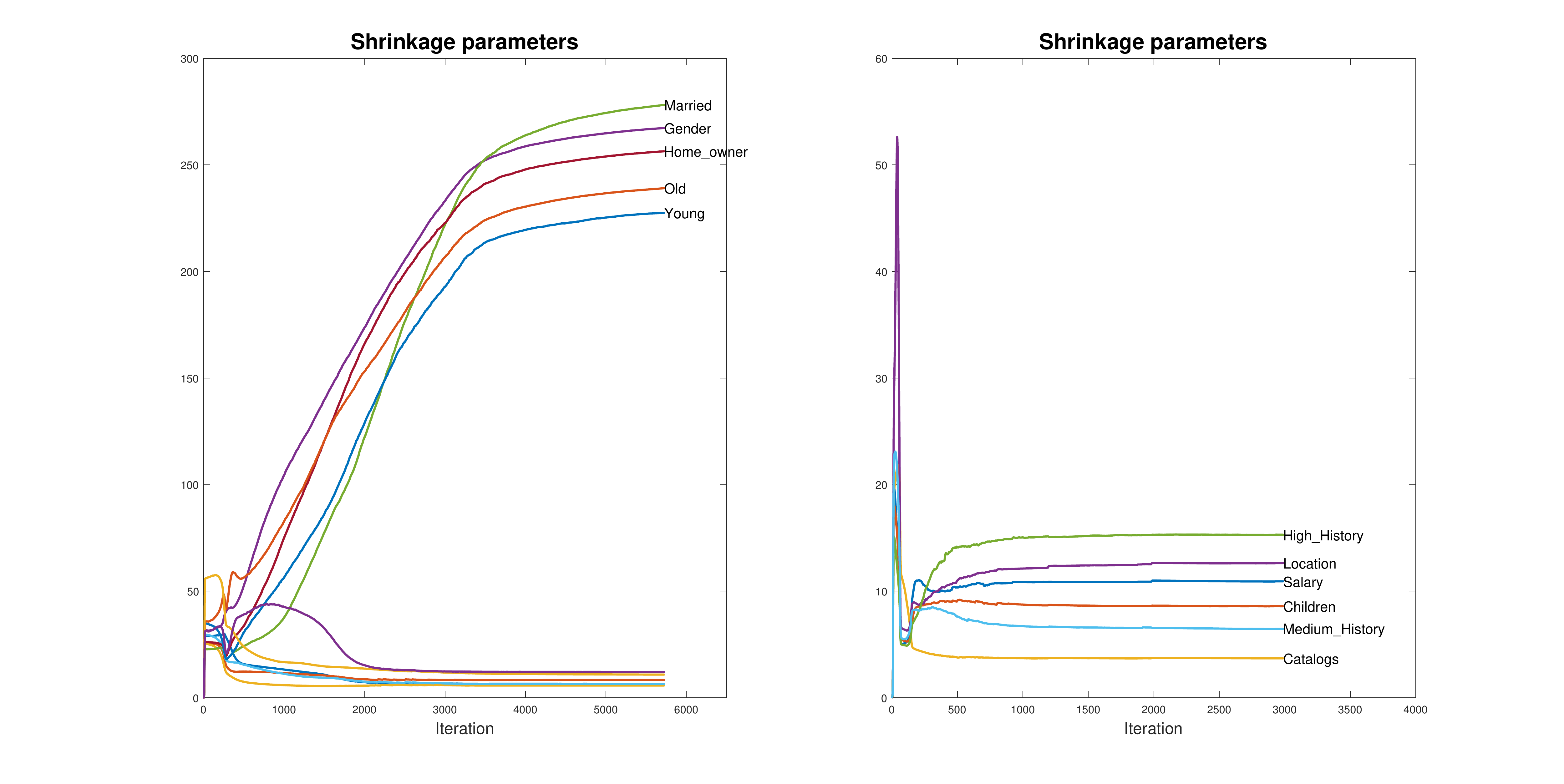}
\caption{Plots of the shrinkage parameters over iterations. Left panel: The shrinkage parameters from a neural net with 6 hidden units. 
Right panel: The shrinkage parameters from a (6,6,6,1) neural net, after the five insignificant covariates have been removed.
}
\label{f:direct marketing: shrinkage}
\end{figure}

We consider the Direct Marketing dataset used in the statistics textbook of \cite{Jank:2011}. This dataset
consists of 1000 observations, of which 900 were used for training and the rest for testing. The response is the amount (in \$1000) a
customer spends on the company's products per year. There are 11 covariates including gender, income, the number of ads catalogs, married status, young, old, etc. The careful
analysis in Jank (2011) shows that the ordinary linear regression model fits well to this dataset. 
We first wish to explore the significance of the covariates in terms of explaining the response $y$.
We tried many DeepGLM models with one hidden layer neural network and the number of units varying,
the plots of the shrinkage parameters over iterations have a consistent pattern and all show that the shrinkage parameters with respect to the covariates married, gender, home owner, old and young increase over iterations.
The left panel of Figure \ref{f:direct marketing: shrinkage} plots the shrinkage parameters from a DeepGLM with 6 hidden units, with an MSE of 0.2325.
The plot suggests that these five covariates can be removed from the model.
The right panel of Figure \ref{f:direct marketing: shrinkage} plots the shrinkage parameters from a DeepGLM with a neural net structure (6,6,6,1), after the five insignificant covariates have been removed.
The MSE for this model is 0.1718. Table \ref{tab:direct marketing} shows that DeepGLM gives a better predictive performance than its competitors.
We note, however, that without considering variable selection, we could not successfully train a DeepGLM model
that has a better predictive performance than BART.
 
\begin{table}[h]
\begin{center}
\begin{tabular}{lll}
\hline\hline
Model&PPS&MSE\\
\hline
GLM&$-0.11$&0.29\\
BART&$-0.35$&0.18\\
DeepGLM&$-0.38$&0.17\\
\hline\hline
\end{tabular}
\end{center}
\caption{Direct marketing data after the five insignificant covariates have been removed: Performance of GLM, BART and DeepGLM. Neural net structure is (6,6,6,1).}\label{tab:direct marketing}
\end{table}

\subsubsection{Abalone data}
The abalone data, available on the UCI Machine Learning Repository, is a benchmark dataset that has been used in many regression analysis papers.
The data has 4177 observations of which 85\% were used for training and the rest for testing.
We first explore the use of the lower bound for model selection. Table \ref{tab:abalone LB and MSE} summarizes  
the lower bound and MSE (computed on the test data), averaged over 10 different runs, for various neural network structures. 
Here, we use the MSE on the test data in order to assess the usefulness of the lower bound as a model selection criterion.  
The results suggest that, in general, a small lower bound leads to a worse MSE and that a DeepGLM with either neural net (9,5,5,1) or (9,10,10,1) can be selected.
We conducted the same experiment for other structures (9,6,6,1), (9,7,7,1), (9,8,8,1) and (9,9,9,1) and observed a little change in both LB and MSE.
This result is consistent with the findings that have been long realized 
in the deep learning literature \citep{Bengio:2012} that a small change around a neural net structure that works well 
does not affect the predictive performance appreciably.
In Table \ref{tab:abalone LB and MSE}, 
the neural nets (9,5,5,1) and (9,10,10,1) have similar LB, but the former should be selected as it has a simpler structure.
This experimental exploration illustrates the attractiveness of using the lower bound as a model selection tool in our NAGVAC method. 
   
\begin{table}[ht]
\begin{center}
\begin{tabular}{l|cccc}
\hline\hline
Structure&$[9,2,2,1]$&$[9,5,5,1]$&$[9,10,10,1]$&$[9,20,20,1]$\\
\hline
LB&$-2.212 (0.019)$&$-2.193 (0.012)$&$-2.190 (0.010)$&$-2.446 (0.060)$\\
MSE&5.17 (0.32)&4.71 (0.12)&4.74 (0.13)&8.61 (1.52)\\
\hline\hline
\end{tabular}
\end{center}
\caption{Abalone data: Lower bound LB and MSE (on the test data), averaged over 10 runs, for various neural network structures. The numbers in brackets are standard errors.}\label{tab:abalone LB and MSE}
\end{table}

\begin{figure}[ht]
\centering
\includegraphics[width=140mm,height=70mm]{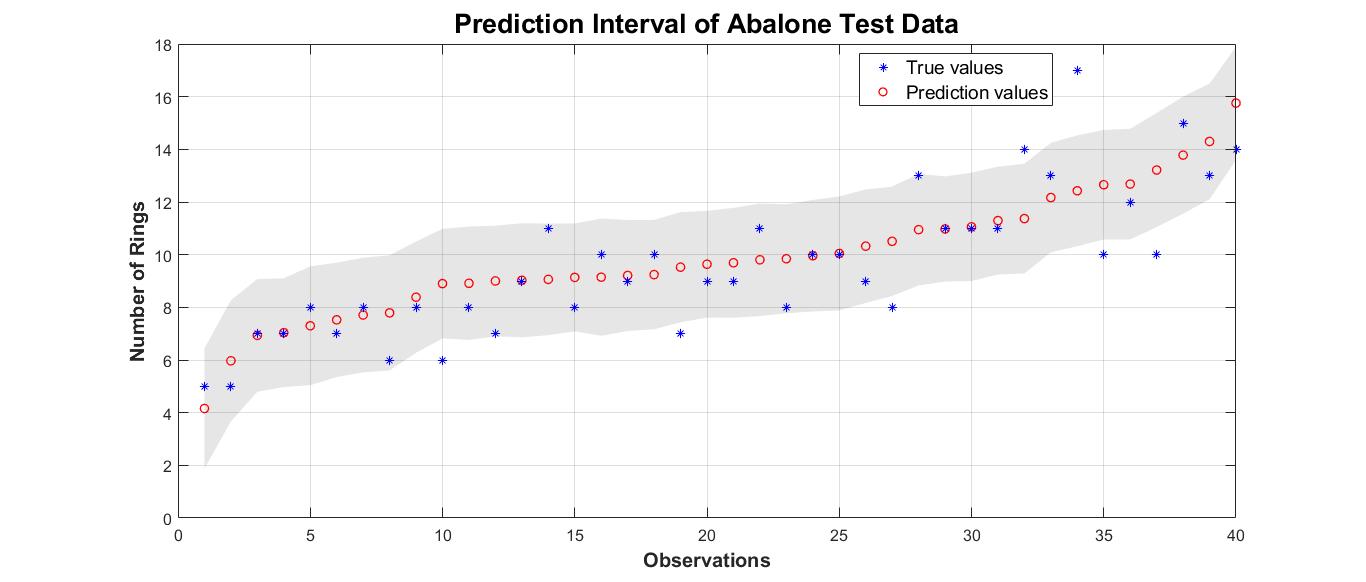}
\caption{Abalone data: Point prediction and one standard deviation prediction intervals (the shaded area) calculated from 
the predictive distribution in \eqref{eq:predictive distribution}.}\label{f:abalone: predictive interval}
\end{figure}

One of the attractive features of DeepGLM is that, as a Bayesian method, it offers an easy and principled way to construct the prediction intervals for test data.
The predictive distribution of the response $y$ given covariate vector $x$ and data $D$ is
\beq\label{eq:predictive distribution}
p(y|x,D)=\int p(y|x,\theta)p(\theta|D)d\theta.
\eeq
As we approximate the posterior $p(\theta|D)$ by the VB distribution $q_\lambda(\theta)$ which we can sample from, it is possible to sample from $p(y|x,D)$ (assuming that it is easy to sample from $p(y|x,\theta)$, which is the case all of our applications).
Based on this sample from the predictive distribution, we can compute prediction intervals for the mean $\E(y|x,D)$.
Figure \ref{f:abalone: predictive interval} shows the one standard deviation prediction intervals for the test data. 

\begin{table}[h]
\begin{center}
\begin{tabular}{lll}
\hline\hline
Model&PPS&MSE\\
\hline
BART&1.30 (0.01)&4.88 (0.06)\\
DeepGLM&1.28 (0.01)&4.71 (0.12)\\
\hline\hline
\end{tabular}
\end{center}
\caption{Abalone data: Performance of BART and DeepGLM. The neural net structure is (9,5,5,1). The values are averaged over 10 runs with the standard errors in brackets.}\label{tab:abalone comparison}
\end{table}

\subsubsection{Census income data}
\begin{figure}[ht]
\centering
\includegraphics[width=140mm,height=70mm]{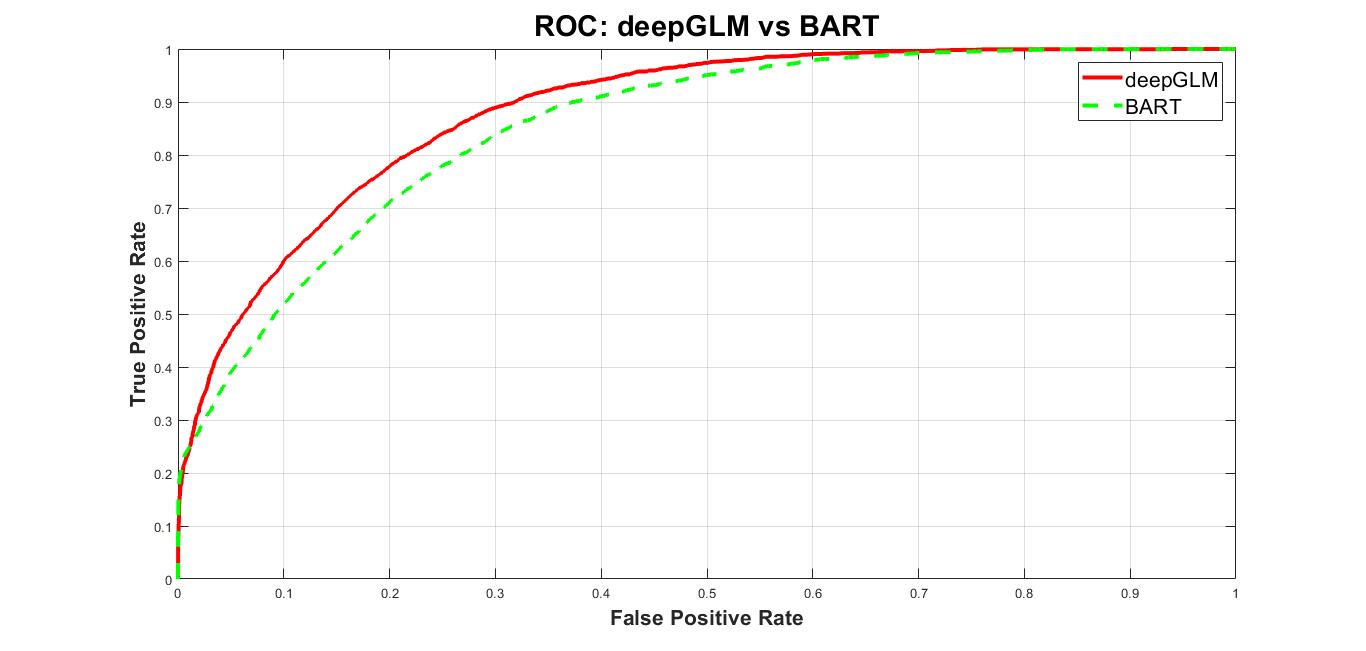}
\caption{Census data: The ROC curves of DeepGLM v.s. BART. The area under the curve of DeepGLM is larger than that of BART, which suggests that 
DeepGLM has a better predictive performance in this example.}\label{f:census: ROC}
\end{figure}

This census dataset was extracted from the U.S. Census Bureau database and is available on the UCI
Machine Learning Repository.
The prediction task is to determine whether a person's income is over \$50K per year,
based on 14 attributes including age, workclass, race, etc, of which many are categorical variables.
After using dummy variables to represent the categorical variables, there are 103 input variables.
There are 45221 observations without missing data, of which 33.3\% are kept for testing,
the rest are used for training.  
Table \ref{tab:census data} summarizes the predictive performance and Figure \ref{f:census: ROC} plots the ROC curves of DeepGLM and BART,
which show that DeepGLM gives slightly better prediction accuracy than BART.
Both DeepGLM and BART are run once with a fixed random seed.

\begin{table}[h]
\begin{center}
\begin{tabular}{lll}
\hline\hline
Model&PPS&MCR (\%)\\
\hline
BART&0.68&18.6\\
DeepGLM&0.34&16.09\\
\hline\hline
\end{tabular}
\end{center}
\caption{Census data: Performance of DeepGLM v.s. BART. We use a one hidden layer neural net with 40 units.}\label{tab:census data}
\end{table}

\subsubsection{A continuous panel data set:  Cornwell and Rupert data}
\label{sec:Cornwell_ Rupert_example}
This section analyzes a continuous panel data set originally analyzed in \cite{Cornwell:88}.  
This is a balanced panel dataset with 595 individuals and 4165 observations,
each individual was observed for 7 years.
The dataset is available from the website of the textbook \cite{Baltagi:2013}.
The variables are listed in Table \ref{tab:Table1}.

\begin{table}[h]
\begin{center}
\begin{tabular}{ll}
\hline\hline
Variable&Meaning\\
\hline
EXP&Years of full-time work experience\\
WKS&Weeks worked\\
OCC&blue-collar occupation = 1; otherwise = 0\\
IND&manufacturing industry=1; otherwise = 0\\
SOUTH &south residence =1; otherwise = 0\\
SMSA&metropolitan residence=1, otherwise = 0\\
MS&married = 1; otherwise = 0\\
FEM &female = 1; otherwise = 0\\
UNION &Union contract wage = 1; otherwise = 0\\
ED & Years of education\\
BLK &black = 1; otherwise = 0\\
LWAGE & log of wage\\
\hline\hline
\end{tabular}
\end{center}
\caption{Cornwell and Rupert data: variables and their meaning}\label{tab:Table1}
\end{table}
We are interested in predicting the wage (on the log scale) of each individual, given the covariates.
Let $y_{it}$ be a continuous variable indicating the log of wage of person $i$ with the vector of covariates $x_{it}$ in year $t$, $t=1,...,7$.
We use the following DeepGLMM model
\beqn
y_{it}|x_{it}\sim\N(\mu_{it},\sigma^{2}),\;\;\;\mu_{it}=\mathfrak{N}(x_{it},w,\beta+\alpha_i),
\eeqn
where $\mathfrak{N}(x_{it},w,\beta+\a_i),\ \beta$ and $\alpha_i$ have the same interpretation as in Section \ref{sec:Binary_panel_data_simulation}, and $\sigma^2$ is the noise variance. 

Since we are interested in within-subject prediction, for each individual, we use the first 5 observations as training data, and the last 2 observations for test data. We use a neural network with 2 hidden layers with 5 nodes each; this structure was selected after some experiments using the lower bound as the model selection criterion.
We compare the performance of DeepGLMM to a linear regression model with a random effect, using PPS and MSE as evaluation metrics.   
Table \ref{tab:schooling_panel_performance} summarizes the results, which show that using the neural network basis functions to model covariate effects in a flexible way can significantly improve both PPS and MSE.

\begin{table}[h]
\begin{center}
\begin{tabular}{lll}
\hline\hline
Model&PPS&MSE\\
\hline
GLMM&0.05&0.18\\
DeepGLMM&-0.87&0.05\\
\hline\hline
\end{tabular}
\end{center}
\caption{Cornwell and Rupert data: Performance of DeepGLMM v.s. conventional GLMM in term of the partial predictive score (PPS) and the mean square error (MSE). Both are evaluated on the test dataset.}\label{tab:schooling_panel_performance}
\end{table}

\subsubsection{Cancer data: high-dimensional logistic regression using the horseshoe prior}
This section illustrates that the training method NAGVAC can be used as a general estimation method for high-dimensional models rather than neural network based models.
The application is concerned with high-dimensional logistic
regression using a sparse signal shrinkage prior, the horseshoe prior \citep{Carvalho2010}.  Here the variational optimization is challenging because of the strong dependence between local variance parameters and the corresponding coefficients.  Using three real datasets we show that the natural gradient estimation method improves the performance of the approach described in \cite{ong+ns16}. 

Let $y_i\in \{0,1\}$ be a binary response with the corresponding covariates $x_i=(x_{i1},\dots, x_{ip})^\top$, $i=1,\dots, n$.  We consider
the logistic regression model
$$\log\frac{\mu_i}{1-\mu_i}=\beta_0+x_i^\top\beta,$$
where $\mu_i=\P(y_i=1|x_i)$, $\beta_0$ is an intercept and $\beta=(\beta_1,\dots, \beta_p)^\top$ are coefficients.  We consider
the setting where $p$ is large, possibly $p\gg n$, and use the horseshoe prior for $\beta$ \citep{Carvalho2010}.  
Specifically we assume $\beta_0\sim N(0,10)$ and 
$$\beta_j|\lambda_j\sim N(0,\lambda_j^2 g^2),\;\;\;\;\;\;\;\;\lambda_j\sim C^+(0,1),\;\;j=1,\dots,p,\;\;\;\;\;\;\;\;g\sim C^+(0,1),$$
where $C^+(0,1)$ denotes the half-Cauchy distribution.  The parameters $\lambda_j$, $j=1,\dots,p,$ are local variance parameters
providing shrinkage for individual coefficients, and the parameter $g$ is a global shrinkage parameter which can adapt to the overall
level of sparsity. 
The above prior settings are the same
as those considered in \cite{ong+ns16}.  

The parameter $\theta$ consists of $\theta=(\beta_0,\beta^\top,\delta^\top,\gamma)^\top$, where $\delta=(\delta_1,\dots, \delta_p)^\top=(\log \lambda_1,\dots, \log \lambda_p)^\top$, and
$\gamma=\log g$.  
We consider Gaussian variational approximation for the posterior of $\theta$, using a factor covariance structure.
The three gene expression datasets are the Colon, Leukaemia and Breast
cancer datasets found at \url{http://www.csie.ntu.edu.tw/~cjlin/libsvmtools/datasets/binary.html}.
These three datasets Colon, Leukaemia and Breast
have training sample sizes of $42$, $38$ and $38$ and test set sizes of $20$, $34$ and $4$ respectively.  
The number of covariates is $p=2000$ for the Colon data, and $p=7120$ for the Leukaemia and Breast datasets.  This means that for 
the Leukaemia and Breast datasets the dimension of $\theta$ is 14,242 so these are examples
with a high dimensional parameter.
These data were also considered in \cite{ong+ns16} where slow convergence in the variational optimization was observed using their method;  
we show here that a natural gradient approach offers a significant improvement.  

\begin{table}[h]
\begin{center}
\begin{tabular}{|c|ccc|ccc|}
\hline
&\multicolumn{3}{c|}{VAFC of \cite{ong+ns16} } & \multicolumn{3}{c|}{NAGVAC-4}\\
\hline
& Train Error &Test Error &CPU &Train Error &Test Error &CPU\\
\hline
Colon &0/42 &0/20 &4.92 &0/42 &0/20 &0.17 \\
Leukemia &0/38 &6/34 &61 &0/38 &1/34 & 0.56\\
Breast &0/38 &1/4 &61.6 &0/38 &0/4 & 0.56 \\
\hline
\end{tabular}
\end{center}
\caption{Performance of the ordinary gradient VAFC and natural gradient VAFC methods on three cancer datasets. Training and test errors rates are reported as the ratio of misclassified data points over the number of data points. Computational time CPU (per 100 iterations) is measured in second.}
\label{tab:High Dim Compare}
\end{table}
Table \ref{tab:High Dim Compare} compares the performance of VAFC of \cite{ong+ns16} and our NAGVAC training methods.
The table shows the predictive performance and computational time on three cancer datasets. We follow \cite{ong+ns16} and run VAFC with $f=4$ factors and use only a single sample to estimate the gradient of lower bound ($S=1$) in two methods. 
As shown, the NAGVAC training method significantly improves the performance of VAFC.

\section{Discussions and conclusions}\label{sec:conclusions}

This paper is concerned with flexible versions of generalized linear and generalized linear mixed models where DFNN methodology is used to automatically
choose transformations of the raw covariates.  The challenges of Bayesian computation are addressed using variational approximation methods 
with a parsimonious factor covariance structure.  We have demonstrated that a natural gradient
approach to the variational optimization with this family of approximations is feasible even in high dimensions. 
Our Bayesian treatment offers a principled and convenient way for selecting the tuning parameters, quantifying uncertainty and doing model selection.   
Using simulated and real datasets and several different models we show that
the improvement that these methods can bring in terms of speed of convergence and computation time are substantial, and that the use of neural network basis functions
with random effects is a class of models that deserve more consideration in the literature.  
 
\section*{Acknowledgments}
Nghia Nguyen and Robert Kohn were partially supported by the Australian Research Council Center of Excellence grant CE140100049.
\section*{Appendix}

\subsection*{Rank-1 natural gradient VB estimation method}
Recall that $\Sigma=BB^\top+D^2$ with $B=(b_1,\cdots,b_d)^\top$ and $D=\diag(c)$, $c=(c_1,\cdots,c_d)^\top$. 
Then,
\[\Sigma^{-1}=D^{-2}-\frac{1}{1+\kappa}D^{-2}BB^\top D^{-2},\;\;\;\kappa_1=B^\top D^{-2}B=\sum\frac{b_i^2}{c_i^2},\]
and $B^\top \Sigma^{-1}B=\kappa_1/(1+\kappa_1)$. Hence, $I_{22}^{-1}=\frac{1+\kappa_1}{2\kappa_1}\Sigma$.
We still set $I_{23}=0$ but $I_{33}$ can be computed analytically as follows.
Let $\Sigma^{-1}=D^{-2}-hh^\top$ with $h=D^{-2}B=(1/\sqrt{1+\kappa_1})B\circ c^{-2}$ and $h^2=h\circ h$.
\bean
I_{33}&=&2D(\Sigma^{-1}\circ \Sigma^{-1})D\\
&=&2\left(D^{-2}-D(hh^\top)\circ D^{-1}-D^{-1}\circ D(hh^\top)+D\big\{(hh^\top)\circ (hh^\top)\big\}D\right)\\
&=&2\left(\diag\big(c^2-2h^2\big)+Dh^2(Dh^2)^\top\right)\\
&=&2\left(\diag(v_1)+v_2v_2^\top\right),
\eean
with $v_1=c^2-2h^2$ and $v_2=Dh^2=c\circ h^2$. Then,
\[I_{33}^{-1}=\frac12\left(\diag(v_1^{-1})+\frac{1}{1+\sum v_{2i}^2/v_{1i}}(v_{1}^{-1}\circ v_2)(v_1^{-1}\circ v_2)^\top\right)\]
It's important to note that it's unnecessary to store the matrix $I_F^{-1}$.
To obtain the natural gradient, all we need is the matrix-vector product of the form $I_F^{-1}g$.
Write $g = (g_1^\top,g_2^\top,g_3^\top)^\top$ with $g_1$ the vector formed by the first $d$ elements of $g$,
$g_2$ the vector formed by the next $d$ elements, and $g_3$ the last $d$ elements in $g$. 
The natural gradient is
\[g^{\text{nat}}=\begin{pmatrix}
(g_1^\top B)B+c^2\circ g_1\\
\frac{1+\kappa_1}{2\kappa_1}\Big((g_2^\top B)B+c^2\circ g_2\Big)\\
\frac12v_1^{-1}\circ g_3+\kappa_2 \big[(v_1^{-1}\circ v_2)^\top g_3\big](v_1^{-1}\circ v_2)
\end{pmatrix},\]
with $\kappa_2=\frac{1}{2}(1+\sum_1^d v_{2i}^2/v_{1i})^{-1}$.
The complexity of computing the natural gradient is $O(d)$.

\subsection*{Further details for the example in Section \ref{sec:Binary_panel_data_simulation}}
The likelihood contribution w.r.t. the panel $(y_i,x_i)$ is
\bean
L_i(\t)&=&\int p(y_i|x_i,w,\b,\a_i)p(\a_i|\Gamma)d\alpha_i\\
&=&\int\exp\left(\sum_{t=1}^{T_i}y_{it}\mathfrak{N}(x_{it},w,\beta+\alpha_i)-\log\big(1+e^{\mathfrak{N}(x_{it},w,\beta+\alpha_i)}\big)\right)p(\a_i|\Gamma)d\a_i.\\
\eean
By Fisher's identity \citep{Gunawan:2017} 
\[\nabla_\t\ell_i(\t)=\int\nabla_\t\left\{\log p(\a_i|\Gamma)+\sum_{t=1}^{T_i}y_{it}\mathfrak{N}(x_{it},w,\beta+\alpha_i)-\log\big(1+e^{\mathfrak{N}(x_{it},w,\beta+\alpha_i)}\big)\right\}p(\a_i|y_i,x_i,\t)d\a_i.\]
We have,
\bean
p(\a_i|y_i,x_i,\t)&\propto& p(\a_i|\Gamma)p(y_i|x_i,w,\b,\a_i)\\
&\propto& \exp\left(\sum_{t=1}^{T_i}\left[y_{it}z_{it}^\top(\beta+\a_i)-\log(1+e^{z_{it}^\top(\beta+\a_i)})\right]-\frac12\a_i^\top\Gamma^{-1}\a_i\right)=\exp(f(\a_i)).
\eean
\[\nabla_{\a_i}f(\a_i)=\v Z_i^\top(y_i-p_i)-\Gamma^{-1}\a_i,\;\;\;p_i=p_i(\a_i)=(p_{i1},...,p_{iT_i})^\top\]
\[\nabla_{\a_i\a_i^\top}f(\a_i)=-\v Z_i^\top\diag(p_i\circ (1-p_i))\v Z_i-\Gamma^{-1}.\]
Let $\wh\mu_{\a_i}$ be the maximizer of $f(\a_i)$ which can be obtained by the Newton-Raphson method, and let 
\beq\label{eq:mu_alpha_i}\wh\Sigma_{\a_i}=\left(-\nabla_{\a_i\a_i^\top}f(\a_i)|_{\a_i=\wh\mu_{\a_i}}\right)^{-1} =\left(\v Z_i^\top\diag(p_i\circ(1-p_i))\v Z_i+\Gamma^{-1} \right)^{-1},\;\;p_i=p_i(\wh\mu_{\a_i})\eeq
We note that for the Gaussian flexible linear mixed model in Section \ref{sec:Cornwell_ Rupert_example}, $\wh\mu_{\a_i}$ and $\wh\Sigma_{\a_i}$ can be derived analytically.

The gradient $\nabla_\t\ell_i(\t)$ can be estimated as follows.
\begin{itemize}
\item Generate $N$ samples $\a_i^{(j)}\sim\N(\wh\mu_{\a_i},\wh\Sigma_{\a_i})$, $j=1,...,N$.
\item Compute the weights
\[w_i^{(j)}=\exp\left(f(\a_i^{(j)})+\frac12(\a_i^{(j)}-\wh\mu_{\a_i})^\top\wh\Sigma_{\a_i}^{-1}(\a_i^{(j)}-\wh\mu_{\a_i})\right)\]
and $W_i^{(j)}=w_i^{(j)}/\sum_{k=1}^N w_i^{(k)}$.
\item Compute the estimate
\[\wh{\nabla_\t\ell_i(\t)}=\sum_{j=1}^N\nabla_\t\left\{\log p(\a_i^{(j)}|\Gamma)+\sum_{t=1}^{T_i}\left[y_{it}z_{it}^\top(\beta+\a_i^{(j)})-\log(1+e^{z_{it}^\top(\beta+\a_i^{(j)})})\right]\right\}W_i^{(j)}.
\]
\end{itemize}

Because the parameters $\Gamma_j$ are positive, a suitable transformation is needed before applying the Gaussian VB approximation.
We use the transformation $\t_{\Gamma_j}=\log(\Gamma_j)$, $j=0,...,M$.
Let $\wt\t=(w,\b,\t_{\Gamma_0},...,\t_{\Gamma_m})$, then, 
\[\t=\t(\wt\t)=\left(w,\b,\exp(\t_{\Gamma_0}),...,\exp(\t_{\Gamma_m})\right).\]
The posterior distribution of $\wt\t$ is 
\[p(\wt\t|D)\propto\left|\frac{\partial\t(\wt\t)}{\partial \wt\t}\right|p(\t(\wt\t))p(\t(\wt\t)|D)= \exp (\t_{\Gamma_0}+...+\t_{\Gamma_m}))p(\t(\wt\t))p(\t(\wt\t)|D)\]
We then approximate $p(\wt\t|D)$ by $q_\l(\wt\t)$.

\bibliographystyle{apalike}
\bibliography{references_v1}

\end{document}